\documentclass[twocolumn,showpacs,preprintnumbers,amsmath,amssymb,superscriptaddress]{revtex4}
\pdfoutput=1

\usepackage{float,graphicx,subfigure}
\usepackage{dcolumn,multirow}
\usepackage{bm}
\usepackage{psfrag}

\begin{document}
\title{Dynamic phase transition and hysteresis dispersion law of the kinetic Ising model with next-nearest neighbor interaction}
\author{William D. Baez}
\affiliation{Department of Chemistry and Physics, Augusta State University, Augusta, GA 30904}
\author{Trinanjan Datta} 
\affiliation{Department of Chemistry and Physics, Augusta State University, Augusta, GA 30904}
\email{tdatta@aug.edu}
\date{\today}
\begin{abstract}\label{abstract}
We study the effects of next-nearest neighbor (NNN) interaction on the dynamic phase transition (DPT) and hysteresis loop area law in the two-dimensional ferromagnetic kinetic Ising model. We find that inclusion of the NNN interaction causes the DPT boundary of the NN kinetic Ising model to shift to larger values of magnetic field and temperature. The NNN kinetic Ising model can therefore exhibit an interaction induced DPT. Also in the low frequency limit (f$\rightarrow$0) the hysteresis loop area law, A(h$_{o}$,f), changes from h$^{0.70}_{o}$f$^{0.36}$ (NN) to h$^{0.14\pm 0.01}_{o}$f$^{0.13\pm0.01}$ (NNN) where h$_o$ is the external field amplitude and f is the frequency. DPT and hysteresis in the kinetic Ising model arises as a competition between the system's metastable lifetime and the time period of the external field. Including the NNN interaction changes the system's metastable lifetime. This causes the DPT and the hysteretic properties to change. We conclude that the systems metastable lifetime is sensitive not only to the lattice size, external field amplitude, and temperature but also to additional interactions present in the system. Furthermore by tracking the probability density P(Q) of the dynamic order parameter Q we conclude that the DPT is of second order. 
\end{abstract}
\pacs{64.60.Ht, 64.60.Qb,75.10.Hk,75.40.Gb,05.40.-a}
\maketitle
\section{Introduction}
The two-dimensional (2D) ferromagnetic nearest-neighbor kinetic Ising (NNKI) model has been used extensively to study non-equilibrium (NEQ) properties. The NNKI model is composed of the Ising Hamiltonian and a term representing an external oscillatory magnetic field ~\cite{PhysRevA.41.4251,PhysRevB.42.856,PhysRevB.43.3373,PhysRevA.42.7471,Chakrabarti:RevModPhys.71.847,Sides:PhysRevLett.81.834}. The presence of an oscillatory field introduces an explicit time dependence in the Hamiltonian. This causes the system to exhibit a hysteretic response which is a characteristic feature of NEQ phenomena. Below the critical temperature (T$_c$) the system exhibits a dynamic phase transition (DPT) ~\cite{Sides:PhysRevLett.81.834,Chakrabarti:RevModPhys.71.847,PhysRevE.56.2407,PhysRevE.56.1234,PhysRevE.58.174,PhysRevE.58.179,PhysRevE.59.218,Sides:PhysRevE.59.2710,Korniss:PhysRevE.63.016120,Korniss:PhysRevE.66.056127} based upon the strength of the field amplitude (h$_0$), frequency (f), and temperature (T) of the system.

Hysteresis in this model occurs due to the competition between the system's metastable lifetime and the time period of the external field ~\cite{Sides:PhysRevLett.81.834}. The hysteresis loop area law for the model in 2D, 3D, and 4D has been investigated using various methods such as Monte Carlo simulations
~\cite{PhysRevA.42.7471,Chakrabarti:RevModPhys.71.847, Sides:PhysRevLett.81.834,Liu:PhysRevB.70.132403,Liu:PhysRevB.65.014416}, mean-field theory ~\cite{PhysRevA.41.4251,PhysRevB.42.856,Luse:PhysRevE.50.224,Liu:PhysRevB.70.132403,Liu:PhysRevB.65.014416} and O(N) type models ~\cite{PhysRevB.43.3373}. Within Monte-Carlo simulations it has been found that the loop area, A(h$_{o}$,f) $\propto$ h$^{0.7}_{o}$f$^{0.36}$ in the low frequency limit f$\rightarrow$0 at a constant temperature ~\cite{Chakrabarti:RevModPhys.71.847}. The scaling relations for the loop area have also been measured in several ultrathin and thin ferromagnetic film systems, such as Fe/Au(001) ~\cite{PhysRevLett.70.2336}, Fe$_{20}$Ni$_{80}$ ~\cite{PhysRevB.60.11906}, Co/Cu(001) ~\cite{PhysRevB.52.14911,PhysRevB.59.4249}, Fe/GaAs(001) ~\cite{PhysRevB.60.10216}, and Fe/W(110)~\cite{PhysRevLett.78.3567}. 

Monte Carlo simulations and mean field studies of hysteresis in this model show that the loop area undergoes a transition from a symmetric shape to an asymmetric shape or vice-versa ~\cite{Chakrabarti:RevModPhys.71.847,Liu:PhysRevB.70.132403,Liu:PhysRevB.65.014416}. This can happen when the driving frequency of the field increases (decreases) beyond a threshold value or when the external field amplitude decreases (increases) for a particular temperature below the critical temperature (T$_c$) of the model. The breakdown in the shape of the hysteresis loop has been attributed to a DPT with a \emph{spontaneously broken symmetric phase} ~\cite{PhysRevA.42.7471,Chakrabarti:RevModPhys.71.847}. A measure of the DPT is given by the dynamic order parameter, Q, which is the period averaged magnetization. The DPT occurs between an ordered dynamic phase with $\langle|Q|\rangle \neq 0$ and a disordered dynamic phase with $\langle|Q|\rangle \approx 0$ ~\cite{Sides:PhysRevLett.81.834}. The DPT in the NNKI model is of second order ~\cite{Sides:PhysRevLett.81.834,Sides:PhysRevE.59.2710,Korniss:PhysRevE.63.016120}. The order of the phase transition was confirmed by tracking the probability density, P(Q), of Q.  

The hysteretic response mechanism and the various phases which arise in the NNKI model is dependent upon the combination of h$_0$, f, and T values. These parameters dictate whether the system will be in one of the four possible regions: strong field (SF), multi-droplet (MD), single-droplet (SD), or co-existence (CE). The SF and the MD regions exist for strong fields/large system sizes and the SD exists for weak fields/small system sizes ~\cite{Sides:PhysRevLett.81.834}. Droplet theory as a mechanism for the NEQ response is valid in the MD and SD regions but breaks down in the SF region ~\cite{Sides:PhysRevE.59.2710}. The results of the NNKI model within the SF region pertaining to the hysteretic properties, dynamic symmetry breaking, and dynamic phase transitions (DPT) are summarized in the review article Ref.~\onlinecite{Chakrabarti:RevModPhys.71.847}. 

In this paper we extend the NNKI model to include the next-nearest neighbor (NNN) interaction. We term this model the next-nearest neighbor kinetic Ising (NNNKI) model (see Eq.~\ref{eq:NNNKI}). We study the effects of NNN interaction on the DPT (see Figs.~\ref{fig:1}-~\ref{fig:3}), correlation (see Fig.~\ref{subfig:Bsurf}), hysteresis loop area law and scaling (see Figs.~\ref{fig:5}~-~\ref{fig:7}), metastable lifetime (see Figs.~\ref{fig:8}~-~\ref{fig:9}), and order parameter fluctuation (see Fig.~\ref{fig:10}). There are two main motivations behind this investigation. First, is the hysteresis loop area law effected by NNN interactions? Second, is the DPT boundary of the NNKI model sensitive to the NNN interactions? 

To obtain an expression for the hysteresis scaling law and to observe the DPT in the NNNKI model we study this model in the low frequency range (f $\in[10^{-3},10^{-2}]$) at a constant temperature T=0.2T$_{c}$, where T$_c$ is the critical temperature of the zero field equilibrium NNN Ising model ~\cite{LandauBinder:PhysRevB.21.1941}. The physical justification for the choice of the frequency range is provided in Section~\ref{sec:hysteresis}. From our study we find that including the NNN interaction changes the hysteresis loop area law from h$^{0.70}_{o}$f$^{0.36}$ (NN) to h$^{0.14\pm 0.01}_{o}$f$^{0.13\pm 0.01}$ (NNN) and shifts the DPT boundary (see Figs.~\ref{subfig:DPT-NNNKI} and ~\ref{fig:2}). The order of the phase transition is second order and this is confirmed by tracking the probability density of Q. We also study the effects of lattice size (L) on the hysteresis loop area law. Furthermore, we explore the scaled variance X$^{Q}_{L}$ of the dynamic order parameter to extract information on the critical properties of the NNNKI model. We also compute the density plot for the correlation function, B, and the hysteresis loop area, A, at a fixed frequency for a range of h$_0$ and T values. 

The reason behind the change in the loop area law and the DPT boundary can be physically understood by exploring the effects of NNN on the metastable lifetime, $\tau$ (see Fig.~\ref{fig:6}). The metastable lifetime is defined as the time taken by the system to decay to a net zero magnetized state (m(t) = 0) from a completely magnetized state (m(t) = ±1) upon an instantaneous field reversal. The sensitive dependence of $\tau$ on L, h$_{o}$ and T has been investigated thoroughly ~\cite{Rikvold:PhysRevE.49.5080}. In this paper we focus on the effect of additional interactions on $\tau$. One of our main findings is that the metastable lifetime is dependent not only on h$_{0}$, T, and L but also on additional interactions present in the model. This change is reflected in the value of the hysteretic exponents and in the location of the DPT phase boundary. Including the interactions however does not change the second order nature of the phase transition. Also, droplet theory as a mechanism for metastable decay still holds in the MD and SD region of the NNNKI model only.

This paper is arranged as follows. In Sec.~\ref{sec:review} we provide a review of the NNKI model. In Sec.~\ref{sec:modelmethod} we state the NNNKI model and describe the Monte Carlo method used to study its properties. In Sec.~\ref{sec:resultsdiscussion} we present and discuss our results on NNNKI model. Finally, in Sec.~\ref{sec:summaryconclusion} we summarize and state the main conclusions of our paper. 
\section{NNKI Review}\label{sec:review} 
Unlike thermally quenched Ising systems the NNKI model is explcitily time-dependent ~\cite{PhysRevLett.71.3222,PhysRevLett.70.3347}. Hysteresis in this model arises out of a competition, R=$\frac{2\pi/\omega}{\langle\tau\rangle}$, between two time scales \cite{Sides:PhysRevLett.81.834}. The quantity R can be thought of as a scaled time or 1/R as a scaled frequency. Here $2\pi/\omega$ is the external field's period and $\tau$ is the system's metastable life. The model's oscillatory nature causes the system to continuously alternate between momentarily stable and unstable states. The metastable state arises when the spins are aligned opposite to the field. When R$<$ 1 the magnetization does not switch signs within a single period and the system has a non-zero value of Q with asymmetric hysteresis loops. However, if R$>$1 a symmetric hysteresis loop is observed and Q=0. 

Metastability has been utilized to show that there exists two decay-mode regimes in the NNKI model - deterministic and stochastic ~\cite{Rikvold:PhysRevE.49.5080}. The deterministic regime is split into the strong field (SF) region and the multidroplet (MD) region. The stochastic regime contains the single droplet (SD) region and the coexistence (CE) region. Within either regime h$_{o}$, L, and T not only determines the value of $\tau$ but also plays a significant role in determining how the system decays from its metastable state.  

Droplet theory provides a physical explanation for the MD and SD regions. Systems within the MD or SD regions experience nucleation and growth of droplets of stable spins over the entire metastable lifetime. Larger systems experiencing MD decay will have many droplets created as it achieves stability. ``Avrami's Law'' provides a classical description of the MD region's mechanics ~\cite{Sides:PhysRevLett.81.834,Sides:PhysRevE.59.2710}. The SF region is characterized by ``droplets'' which are as large as the system size. In this region droplet theory breaks down. The stochastic regime is characterized by weak external fields, small system sizes, and extremely long average metastable lifetimes ($\left\langle \tau(h_{0})\right\rangle > 10^{3}$) ~\cite{Sides:PhysRevE.57.6512}. The system within this region experiences a slow metastable decay as a single droplet of spin nucleates and eventually grows overtaking the system. 

The MD and SD regions have been studied exhaustively for finite size scaling effects, DPT, universality, and critical exponents ~\cite{Sides:PhysRevE.59.2710,Korniss:PhysRevE.63.016120,Korniss:PhysRevE.66.056127}. It has been found that the critical exponents of the NNKI model are close to the equilibrium 2D Ising model and the 2D random percolation model ~\cite{Korniss:PhysRevE.63.016120}. The dynamical response of the NNKI model has also been studied subject to a square-wave oscillating external field with a soft Glauber dynamics ~\cite{buenda:051108,PhysRevLett.92.015701}. 

In Ref.~\onlinecite{Chakrabarti:RevModPhys.71.847} the NNKI models hysteretic response has been studied using dynamic symmetry breaking. The hysteresis loop becoming asymmetric around the origin was used as a signature of DPT ~\cite{Chakrabarti:RevModPhys.71.847}. Rikvold and collaborators used the metastability mechanism to show that a DPT within the NNKI model exists only when crossing between the MD and SF regions. Recently, evidence for a dynamic phase transition has been investigated experimentally in [Co/Pt]$_3$ magnetic multilayers ~\cite{robb134422}. 
\section{Model and Method}\label{sec:modelmethod}
The model used in this study is the kinetic NNN Ising ferromagnet on a square lattice with periodic boundary conditions. The NNNKI Hamiltonian is given by \begin{eqnarray}\label{eq:NNNKI}
H=-J_{nn}\sum_{\langle i,j\rangle}S_{i}S_{j}-J_{nnn}\sum_{[i,j]}S_{i}S_{j}-h_{o}\sin(\omega t)\sum_{i}S_{i}\nonumber\\
\end{eqnarray}where S$_{i}$ is the $i$th spin and can have values of S$_{i}=\pm$ 1, J$_{nn}$ is the NN coupling, J$_{nnn}$ is the NNN coupling, and h$_o$ is the external field amplitude. The sums $\sum_{\langle i,j\rangle}$ and $\sum_{[i,j]}$ run over all the lattice sites, L$^{2}$, for the NN and NNN interaction. Both the couplings are ferromagnetic and J$_{nn}$=J$_{nnn}$=1. The ratio of the couplings is defined as p=J$_{nnn}$/J$_{nn}$=1.

The spin-flip dynamics used is the Metropolis algorithm with the Monte Carlo step per spin (MCSS) as the unit time step ~\cite{LandauBinderbook,NewmanBarkemabook,Sethnabook,GouldTobochnikbook}. The system is allowed to be in contact with a heat bath at temperature T, and each attempted spin flip from S$_{i}\rightarrow$ -S$_{i}$ is accepted with the probability W(S$_{i}\rightarrow$ -S$_{i}$)=$\exp(-\beta \Delta$ E$_{i}$). Here $\Delta E_{i}$ is the change in energy of the system that would result if the spin flip were accepted, and $\beta$=1/k$_{B}$T where the Boltzmann constant k$_{B}$ is set equal to one. 

Using the above Hamiltonian (Eq.~\ref{eq:NNNKI}) and the Monte Carlo method we compute Q, the correlation (B), and the area (A) using
\begin{eqnarray}
{\label{eq:Qform}Q=\frac{\omega}{2\pi}\oint m(t) dt},\\
{B=\frac{\omega}{2\pi}\oint m(t)h(t)dt\label{eq:Bform}}, \\
{A=\oint m(h)dh\label{eq:Aform}}.
\end{eqnarray} 
where m(t) is the time dependent magnetization per unit site. In the next two sections we study the physics of the NNKI model with NNN interactions.
\section{Results and Discussion}\label{sec:resultsdiscussion}
In this section we report results from the Monte Carlo simulation of the NNNKI model. We calculate the DPT plot, P(Q) graphs, hystersis loop area exponents and scaling, correlation, and fluctuation of Q. The MCSS value used for these computations is 3.5$\times$10$^{5}$. Furthermore, in this paper we focus on the deterministic (SF/MD) region in the low frequency regime (f $\in$ 10$^{-3}$ to 10$^{-2}$). 
\begin{figure}[h]
\centering
\subfigure[Dynamic phase transition plot]{\includegraphics[width=3.5in]{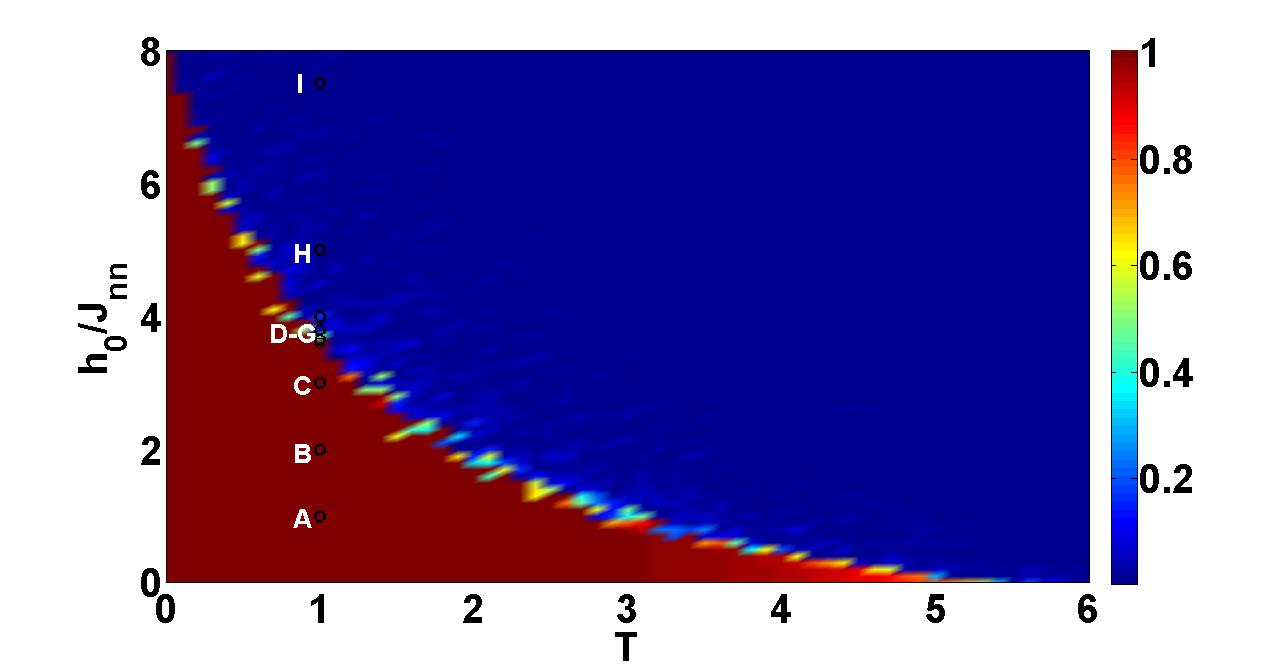}\label{subfig:DPT-NNNKI}}
\subfigure[Probability density, P(Q)]{\includegraphics[width=3.5in]{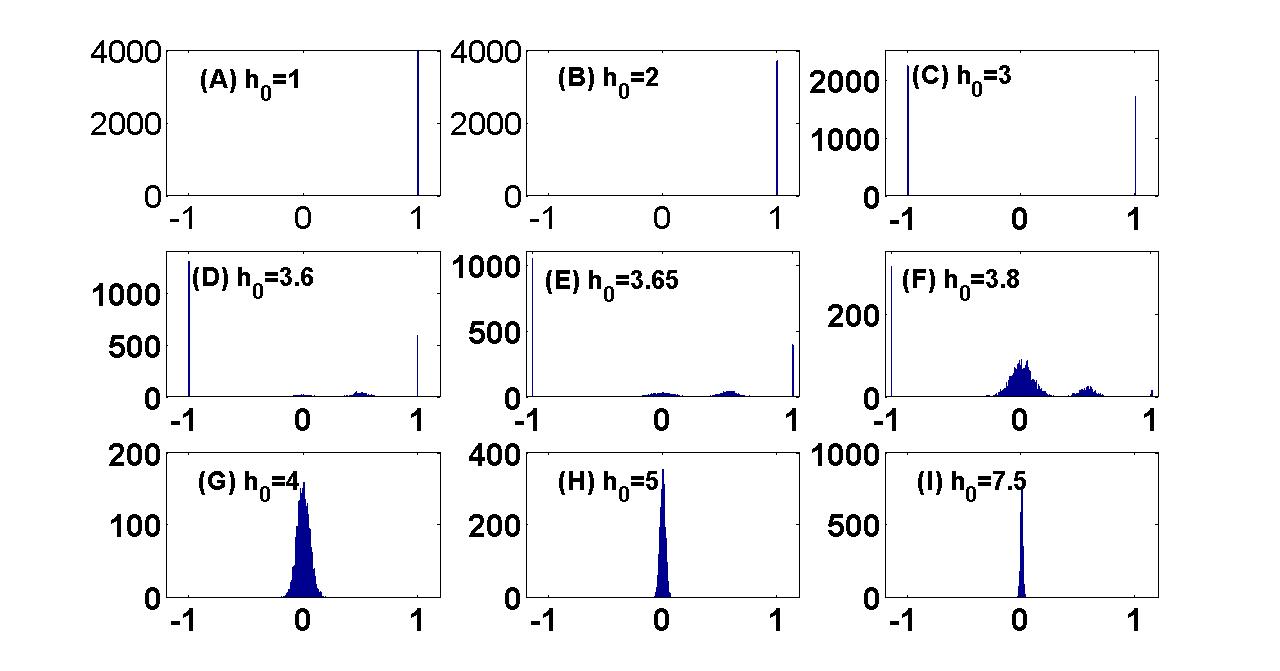}\label{subfig:PQ}}
\caption{(Color online) Plots of the dynamic phase transition (DPT) and the probability density, P(Q), of the dynamic order parameter Q for L=64 at f=10$^{-3}$. h$_{0}$ is the external field amplitude, T is the temperature, and the NN interaction is J$_{nn}$=1. The letters A~-~I track the DPT along T=0.2T$_{c}$. (a) The overall qualitative feature of the DPT phase diagram for the NNN case is similar to the NN case (see Fig.~16c of Ref.~\onlinecite{Korniss:PhysRevE.63.016120} and Fig.~\ref{fig:2} in this paper). The various phases: strong field (SF), multi-droplet (MD), and single droplet (SD) still survive. Quantitatively inclusion of the NNN interaction shifts the DPT phase boundary to higher values of magnetic field and temperature. The presence of additional interactions shift the DPT phase boundary and can give rise to \emph{an interaction induced DPT}. (b) P(Q) plots for values of h$_{0}$ corresponding to the letters A~-~I marked on the DPT plot. The x-axis represents Q and the y-axis P(Q). In the SD region there is a peak at $Q$=1 and in the SF region there is a peak at $Q$=0. In the intermediate MD region there is a double peak. The observation that the P(Q) displays no peak near Q=0 in the ordered dynamic phase is evidence that the DPT is a second order phase transition and it survives upto the NNN interaction.}
\label{fig:1}
\end{figure}
\begin{figure}[h]
{\label{subfig:hyst1a}
  \includegraphics[width=3.5in]{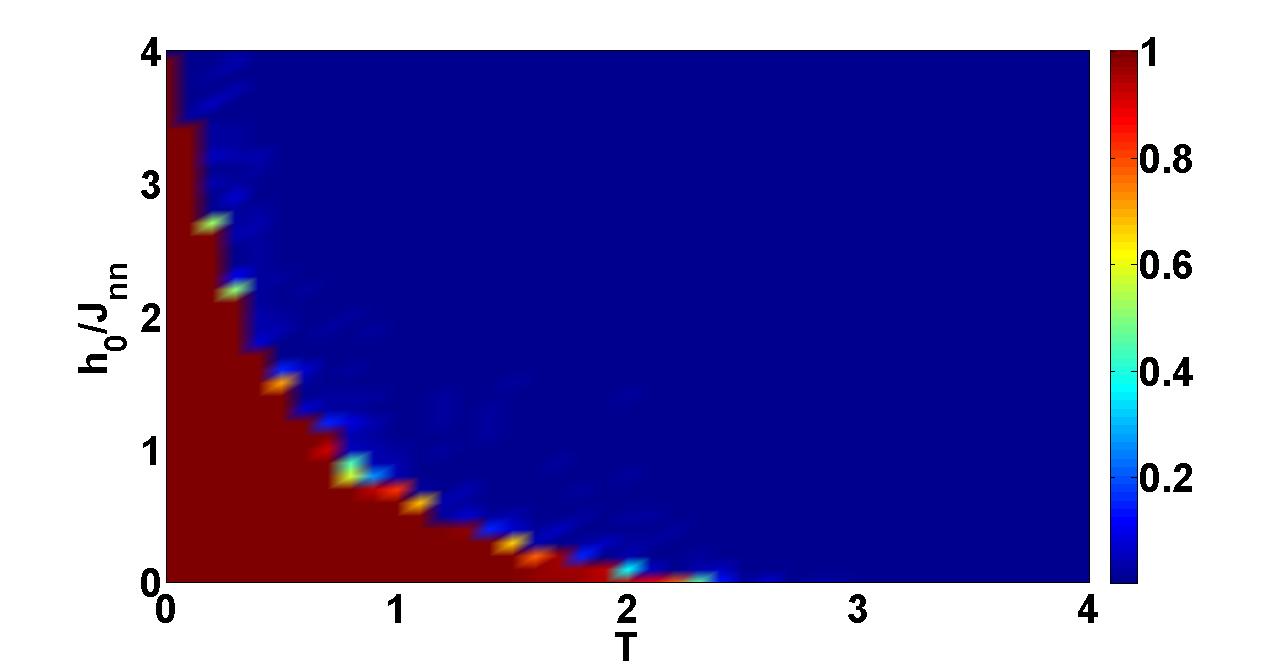}}
  \caption{(Color online) Dynamic phase transition plot for the NNKI model for L=64 and f=10$^{-3}$. h$_{0}$ is the external field amplitude, T is the temperature, and the NN interaction is J$_{nn}$=1. Compared to the NNNKI DPT plot (see Fig.~\ref{fig:1}) the SF/MD region occupies a much smaller space in the phase diagram. \label{fig:2}}
\end{figure}
\begin{figure}[h]
  {\includegraphics[width=3.5in]{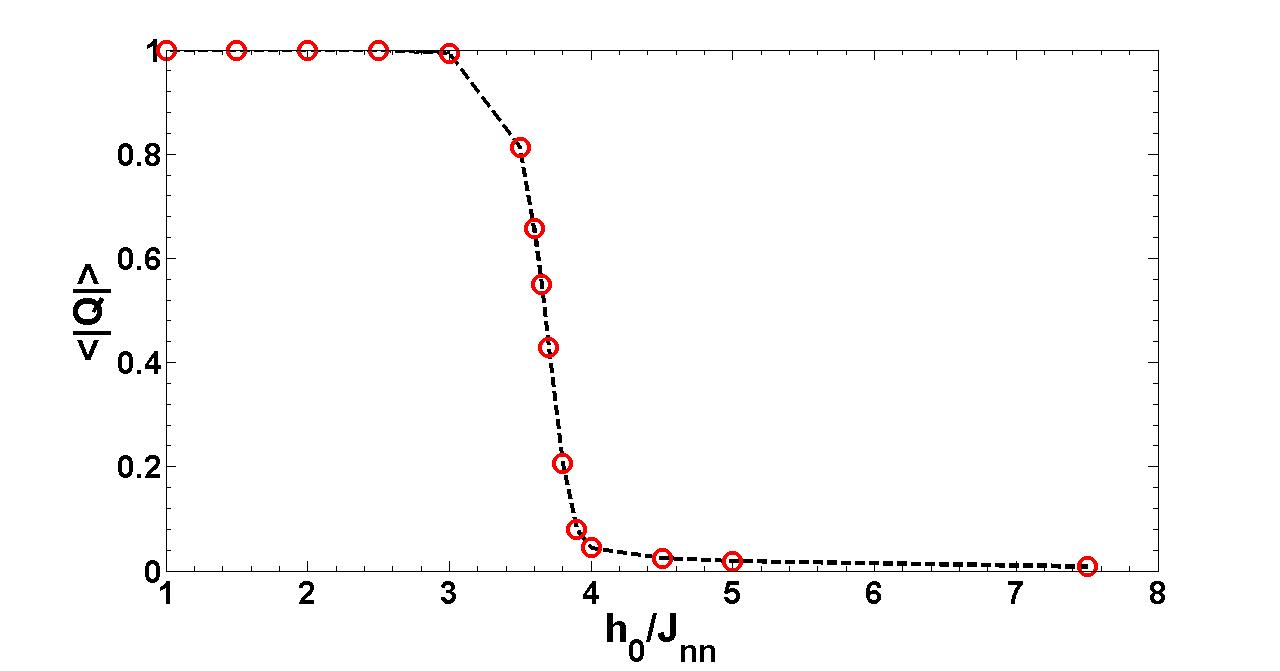}}
  \caption{(Color online) Mean of the norm of the dynamic order parameter $\langle |Q|\rangle$ versus h$_{0}/$J$_{nn}$ corresponding to the letters A~-~I on the DPT plot of Fig.~\ref{fig:1}. h$_{0}$ is the external field amplitude and the NN interaction is J$_{nn}$=1. The values demonstrate the DPT from the SD to the MD and then finally to the SF region. The lines connecting the data points are a guide to the eye. \label{fig:3}}
\end{figure}
\begin{figure}[h]
\begin{center}
\subfigure[Correlation (B)]{\includegraphics[width=3.5in]{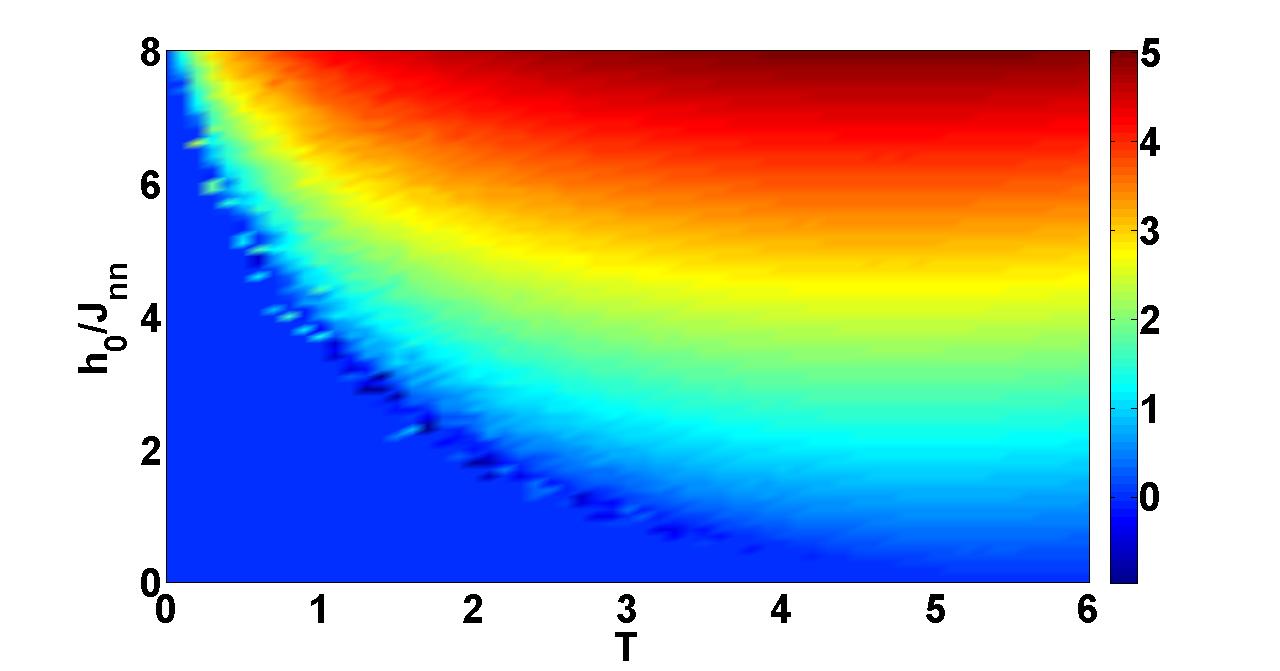}\label{subfig:Bsurf}}
\subfigure[Area (A)]{\includegraphics[width=3.5in]{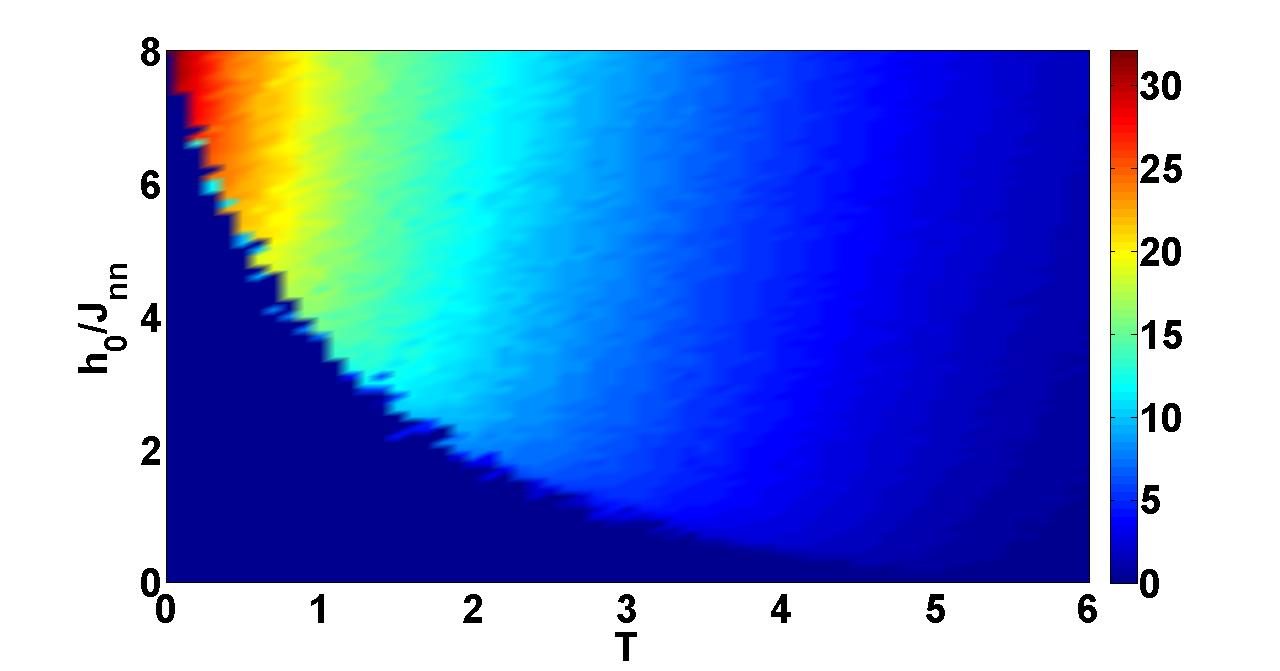}\label{subfig:Asurf}}
\end{center}
\caption{(Color online) Density plots for the correlation, B, and the hysteresis loop area, A, at T=0.2T$_c$ for L=64, f=10$^{-3}$. The external field amplitude is given by h$_{0}$, temperature by T, and the NN interaction by J$_{nn}$, where J$_{nn}$=1. (a) Density plot of the correlation. The system is maximally correlated in the SF region and minimally in the SD region. (b)Density plot for the hysteresis loop area. The system has the largest area for low temperatures and high magnetic fields.}
\label{fig:4}
\end{figure}
\subsection{DPT and Probability distribution, P(Q)\label{sec:pq}}
Fig.~\ref{subfig:DPT-NNNKI} shows the DPT plot for the NNNKI model. The plot displays the values of $\langle |Q|\rangle$ for a range of h$_{o}$/J$_{nn}$ and T for a fixed frequency, f=10$^{-3}$. The figure displays three regions - strong field ($\langle |Q|\rangle$=0), multi-droplet (0$<\langle |Q|\rangle<$1), and single-droplet ($\langle |Q|\rangle$=1). The corresponding DPT plot for the NNKI model is shown in Fig.~\ref{fig:2}. Comparing the two plots we observe that while the overall shape is the same, the boundary of the DPT changes with the inclusion of NNN interactions. For the same choice of temperature the SD/MD region in the NNKI plot occupies a much smaller region compared to the NNNKI-DPT plot. \emph{The system can undergo an interaction induced DPT.} The sensitivity of the DPT phase boundary can be physically explained by considering the effect of interactions on the metastable lifetime of the system (see Section~\ref{sec:tau}). 
\begin{figure}[t]
   \subfigure[L=64]{\label{subfig:area1}
   \includegraphics[width=3.5in]{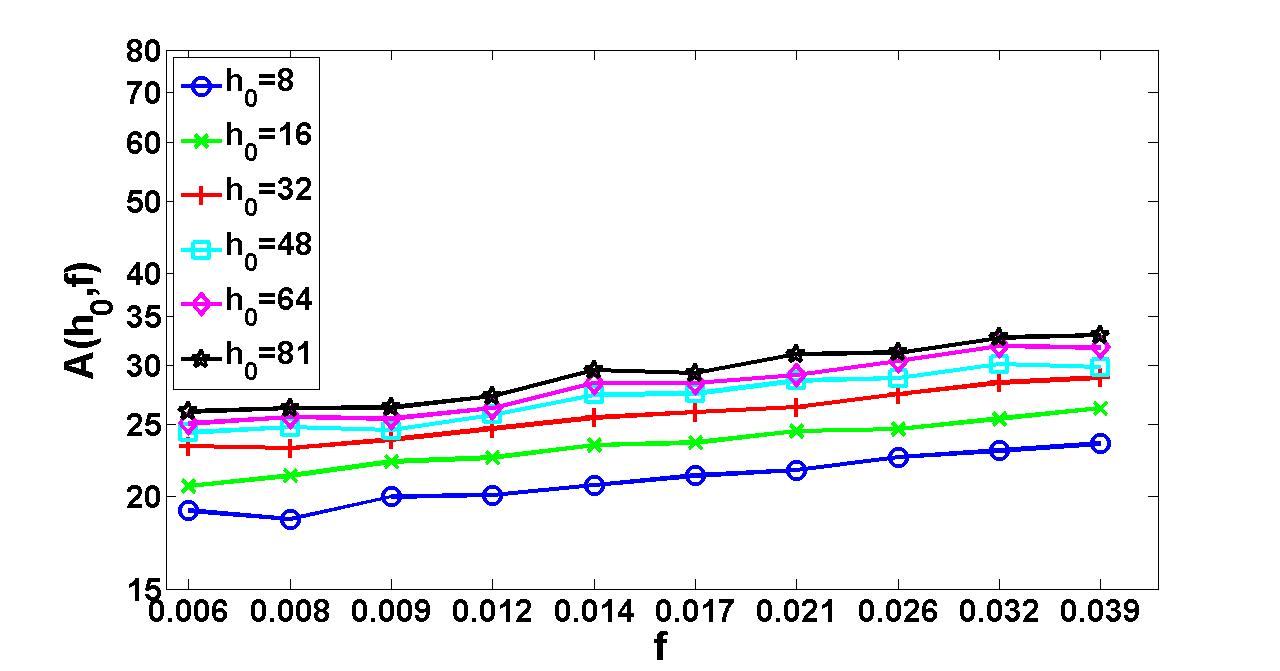}\label{subfig:Area64}}
  \subfigure[L=128]{\label{subfig:area2}
  \includegraphics[width=3.5in]{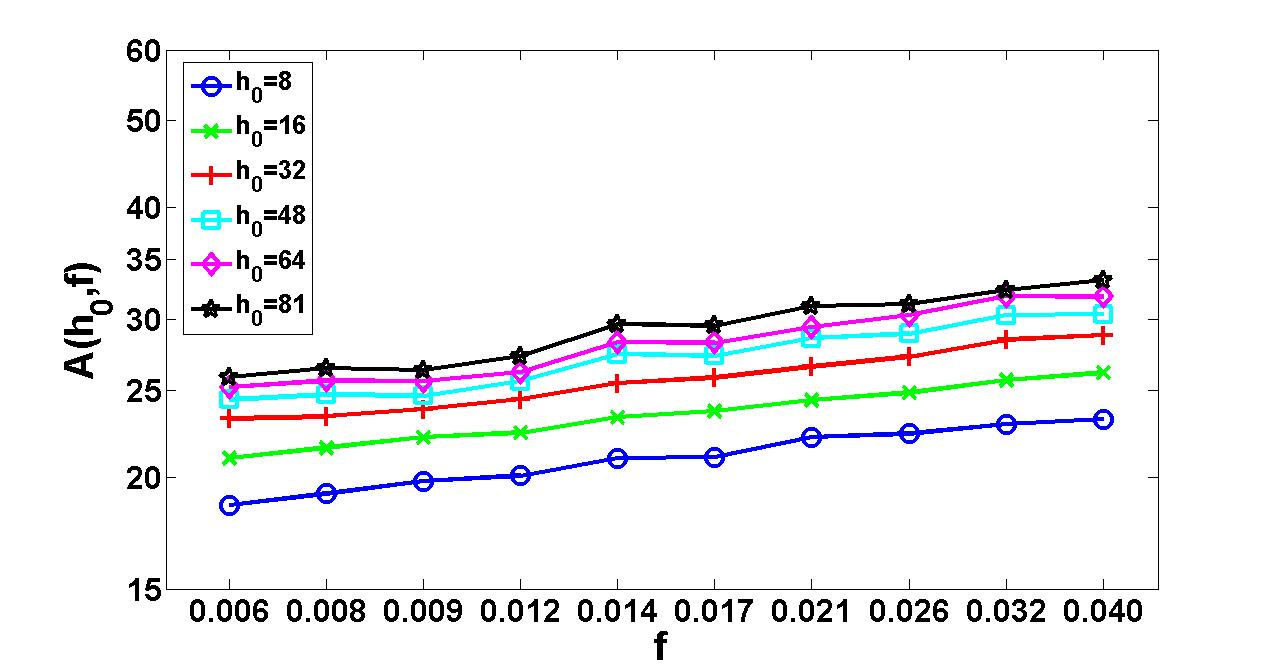}\label{subfig:Area128}}
\caption{(Color online) Log-Log plot of the frequency dispersion of the hysteresis loop area law, A(h$_{o}$,f), for the low frequency and the high magnetic field amplitude regime (h$_{o}\geq$ 8) of the NNNKI model at a temperature T=0.2T$_c$. f $\in[10^{-3},10^{-2}]$ represents the frequency range and h$_{0}$ is the field amplitude. The choice of h$_{0}$ is such that the model is in the SF region. Inclusion of the NNN interaction causes a change in the hysteresis exponents. A(h$_{o}$,f) is modified from h$^{0.70}_{o}$f$^{0.36}$ (NN) to h$^{0.14\pm 0.01}_{o}$f$^{0.13\pm0.01}$ (NNN). The hysteresis loops corresponding to the area variation are shown in Fig.~\ref{fig:4}. The plots are shown for two lattice sizes. The lines connecting the data points are a guide to the eye. (a)L=64 and (b)L=128}
\label{fig:5}
\end{figure}
\begin{figure}[t]
  \subfigure[Hysteresis plot, L=64,h$_{0}$=16]{\label{subfig:hyst1a}
  \includegraphics[width=3.5in]{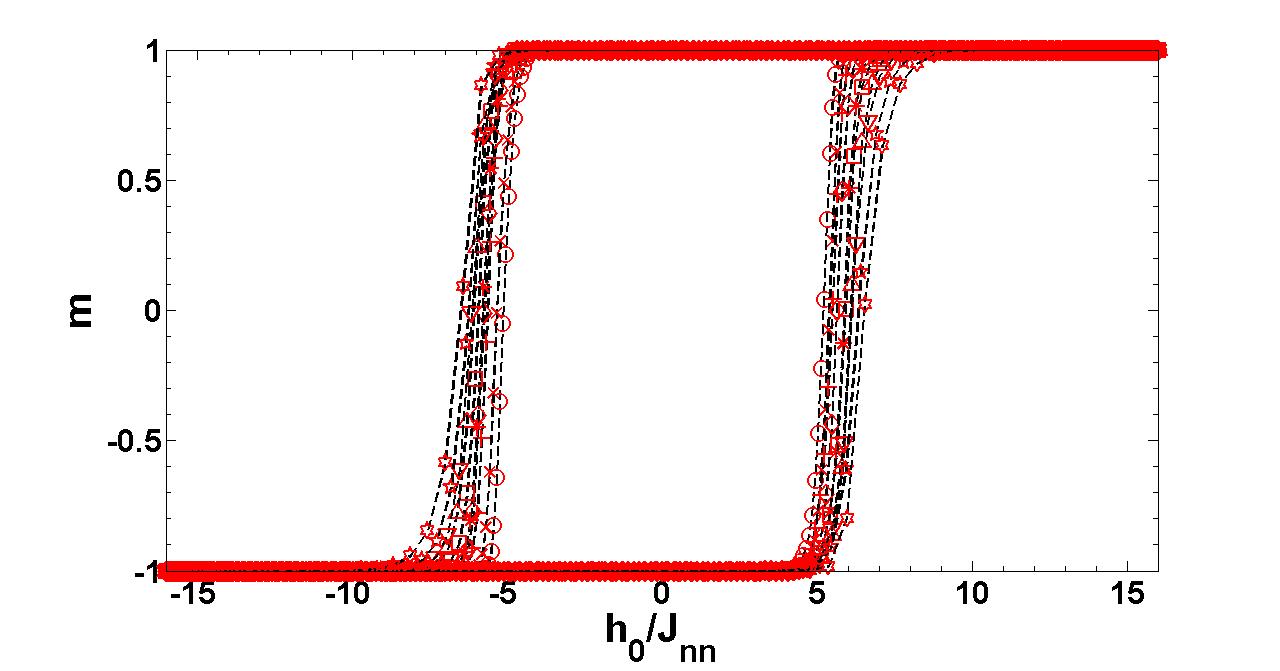}\label{subfig:Hyst64}}
  \subfigure[Hysteresis plot, L=128,h$_0$=16]{\label{subfig:hyst1b}
  \includegraphics[width=3.5in]{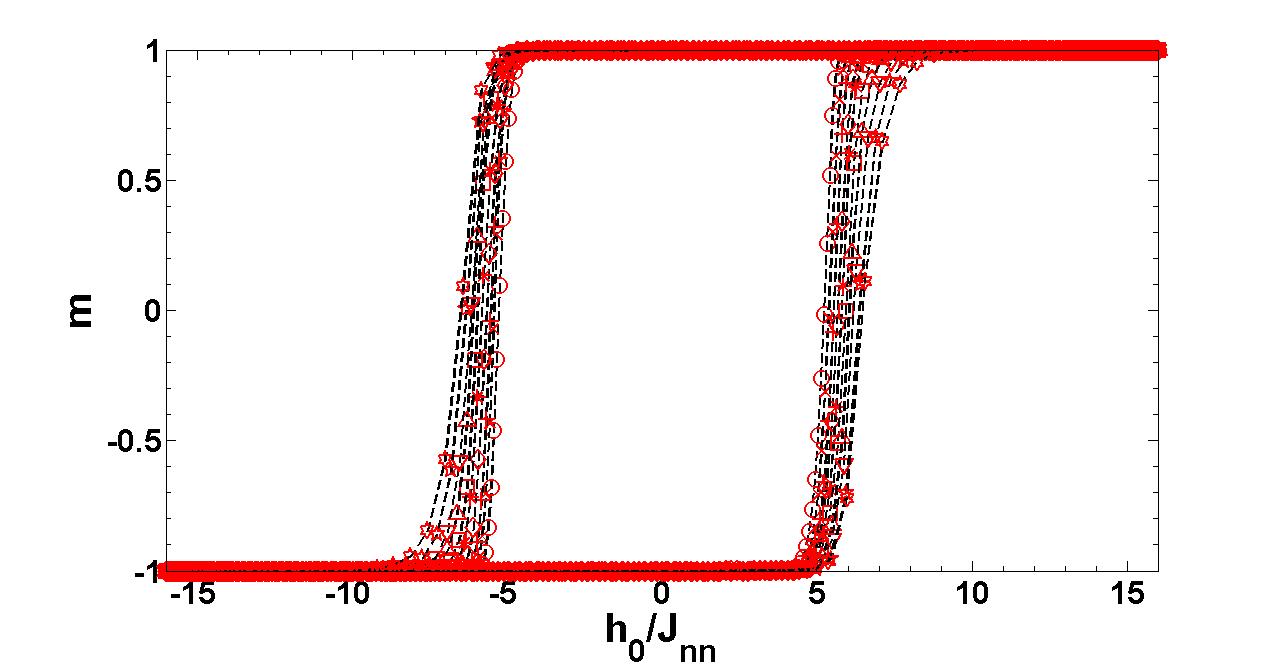}\label{subfig:Hyst128}}
  \caption{(Color online) Hysteresis loops at T=0.2T$_{c}$ and for h$_{0}$=16 for various frequencies. f $\in[10^{-3},10^{-2}]$ represents the frequency range and h$_{0}$ is the field amplitude. The smallest loop area corresponds to f=10$^{-3}$. The plots are shown for two lattice sizes. The dashed lines connecting the data points are a guide to the eye. (a)L=64 and (b)L=128}\label{fig:6}
\end{figure}
\begin{figure}[h]
\centering
\subfigure[Scaling law for loop area (h$_{0}$ variation)]{\includegraphics[width=3.5in]{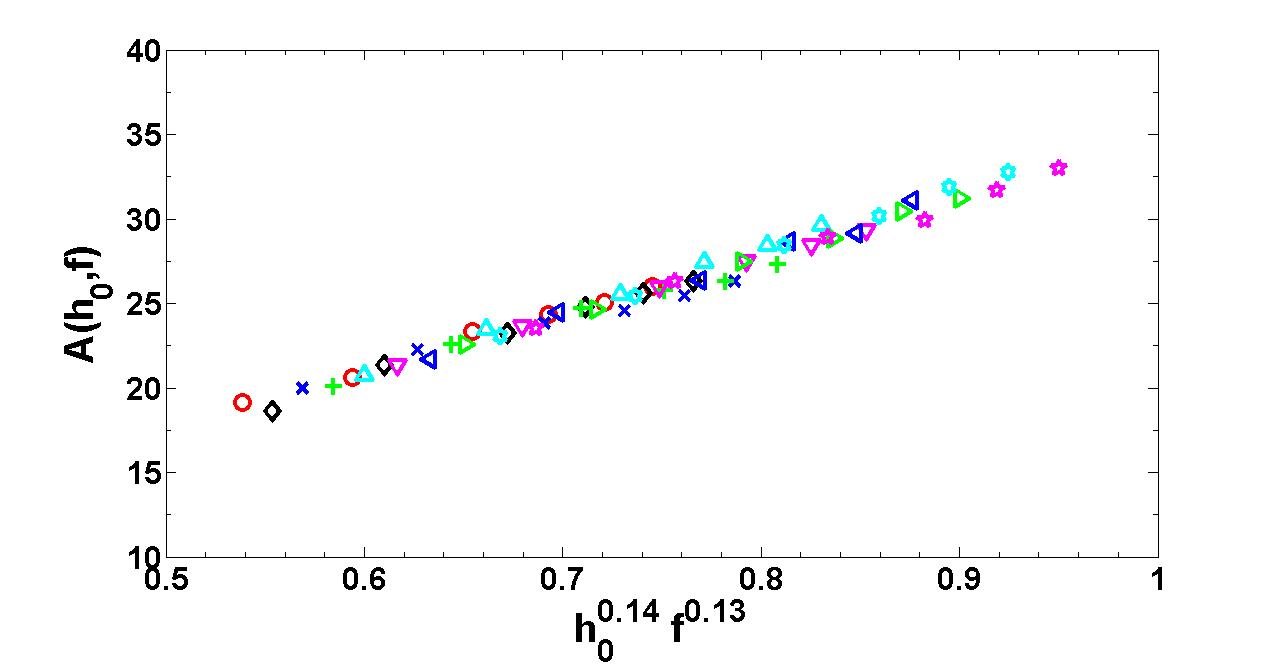}\label{subfig:hscaling}} \subfigure[Scaling law for loop area (frequency variation)]{\includegraphics[width=3.5in]{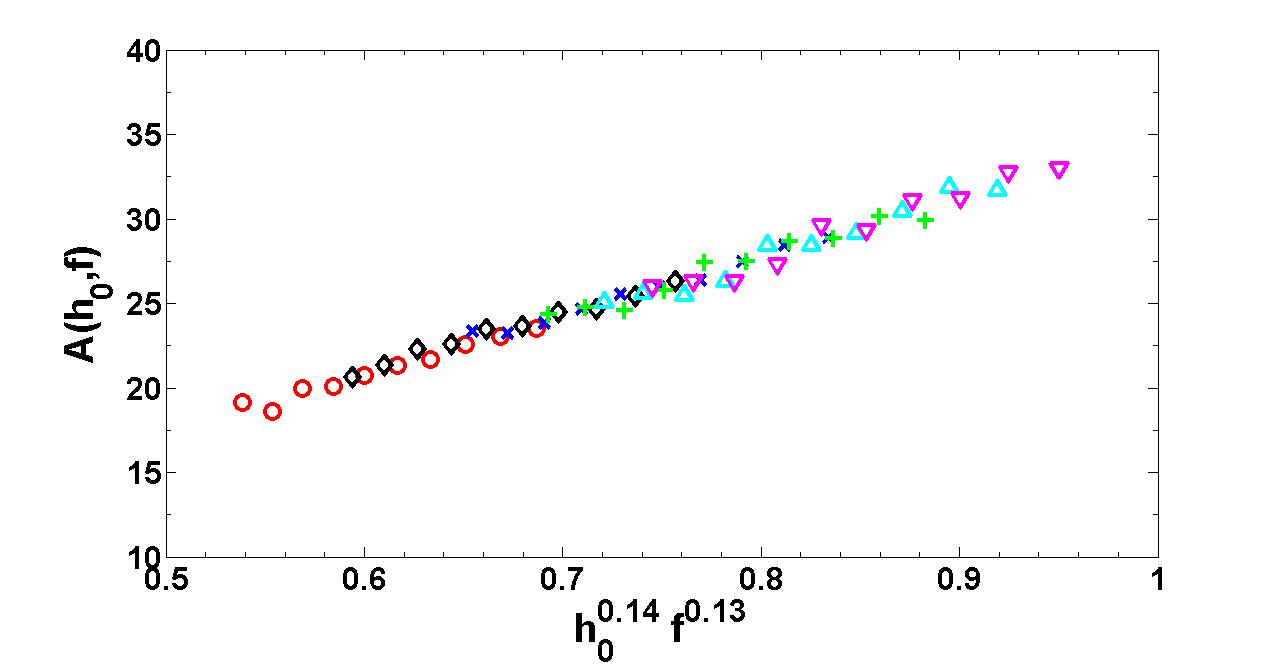}\label{subfig:wscaling}}
\caption{(Color online) Scaling laws for the loop area A(h$_{0}$,f) at a constant temperature T=0.2T$_{c}$. The hysteresis area law data of Fig.~\ref{fig:5} is replotted as function of the scaling variable h$^{0.14}_{0}$f$^{0.13}$. For the frequency range (f $\in[10^{-3},10^{-2}]$) and the external field amplitude range h$_{o}$=8~-~81 considered in this model the collapse of the data is good and the proposed area law is well obeyed. (a) Plot markers correspond to the various values of h$_{0}$=8,16,32,48,64, and 81 (b)The plot markers correspond to the frequency range f $\in[10^{-3},10^{-2}]$.}
\label{fig:7}
\end{figure}
\begin{figure}[b]
{\label{subfig:hyst1a}
  \includegraphics[width=3.5in]{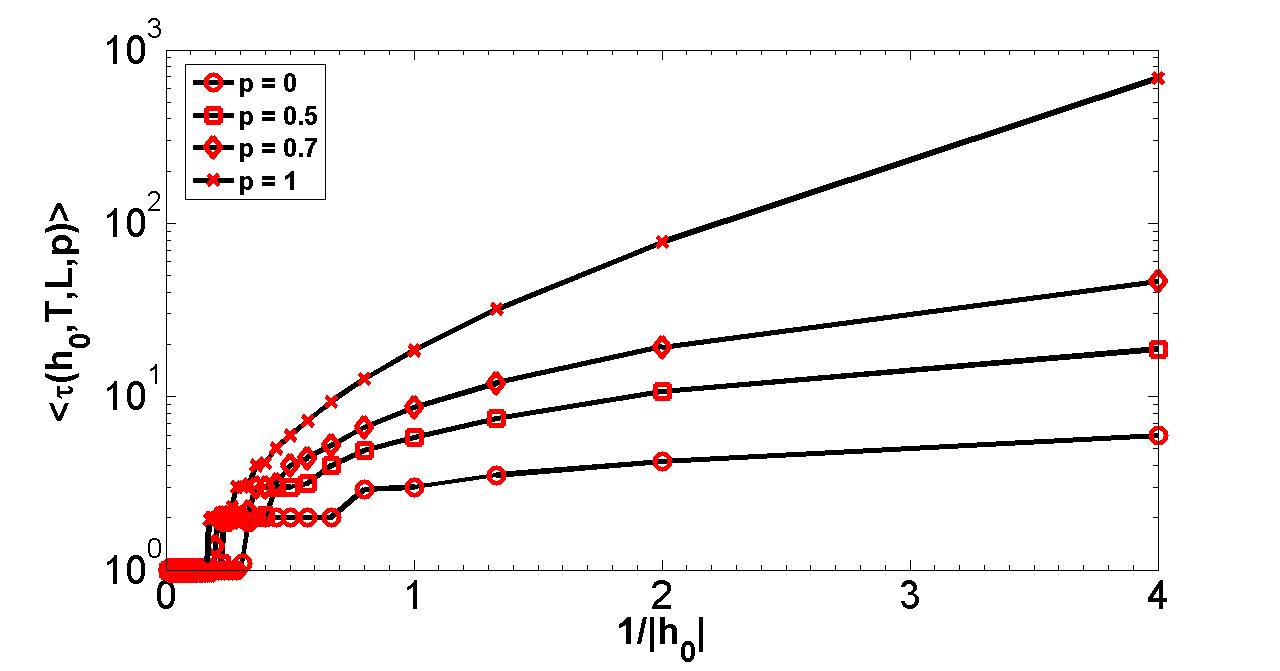}}
  \caption{(Color online) Variation of the average metastable lifetime, $\langle\tau(h_{0},T,L,p)\rangle$ as a function of interaction strength ratio, p=J$_{nnn}$/J$_{nn}$. The calculations were performed at T=0.8T$_{c}$ and h$_{0}$=0.5J$_{nn}$, where J$_{nn}$=1.We see from the plot that the value of $\tau$ is sensitive to additional interactions present in the system. The physics of the NNKI and the NNNKI model is governed by the metastable lifetime. A change in the value of $\tau$ causes the system to produce a shift in the DPT phase boundary and to obey a new hysteresis loop area law. The lines connecting the data points are a guide to the eye.}\label{fig:8}
\end{figure}
\begin{figure}[t]
\begin{center}$
\begin{array}{ccc}
{\includegraphics[width=1.1in]{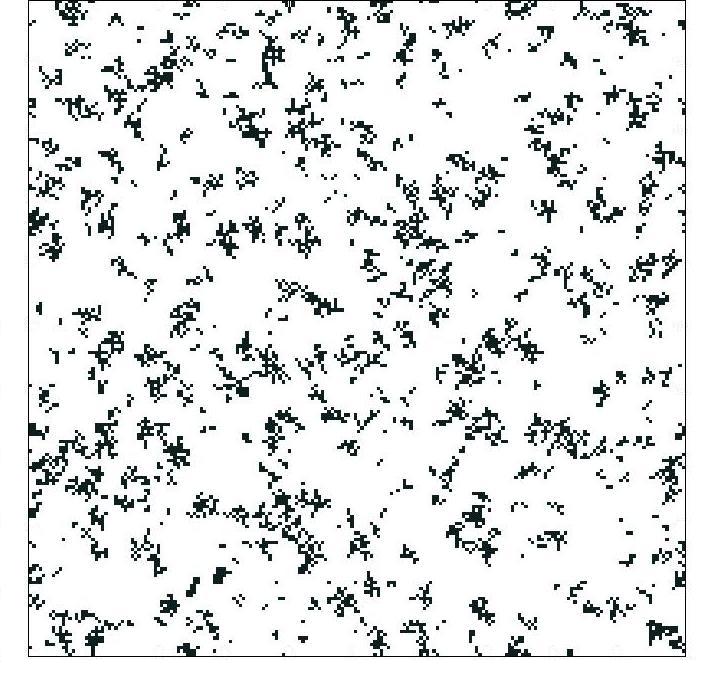}\label{fig:SF1}} &{\includegraphics[width=1.1in]{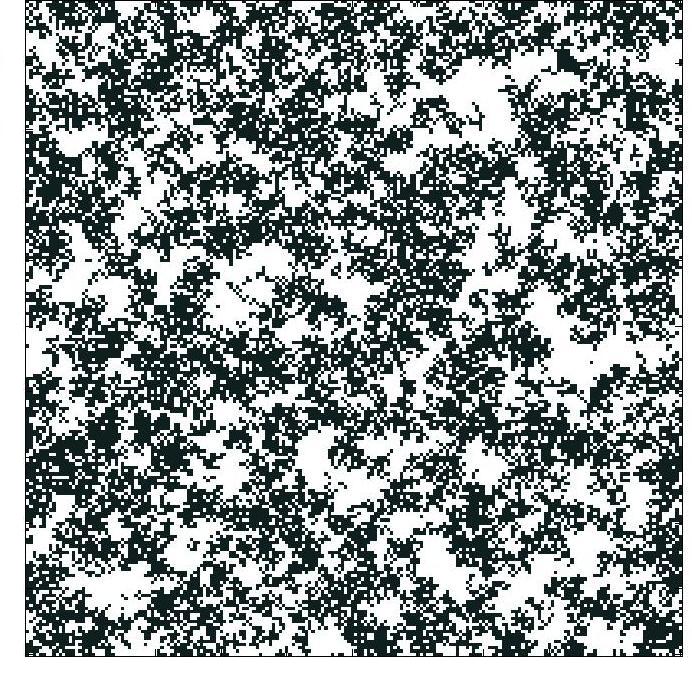}\label{fig:SF2}}
&{\includegraphics[width=1.1in]{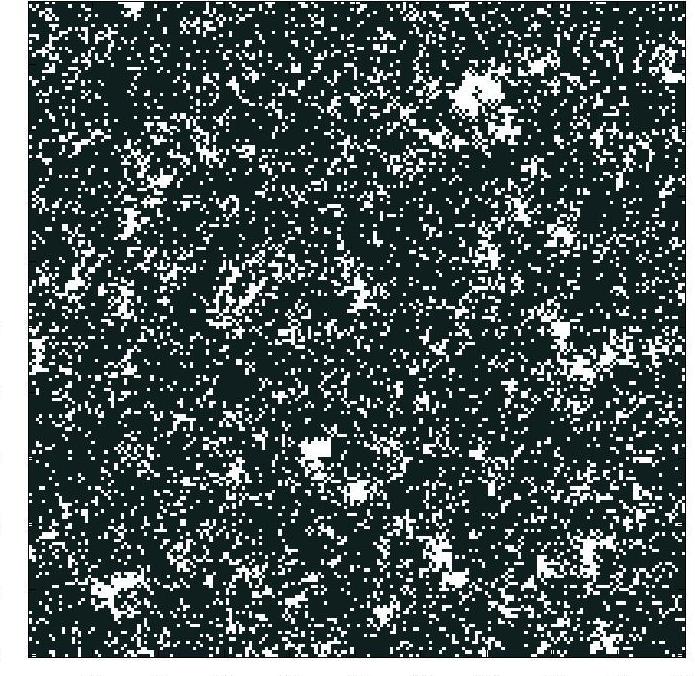}\label{fig:SF3}}\\
{\includegraphics[width=1.0in]{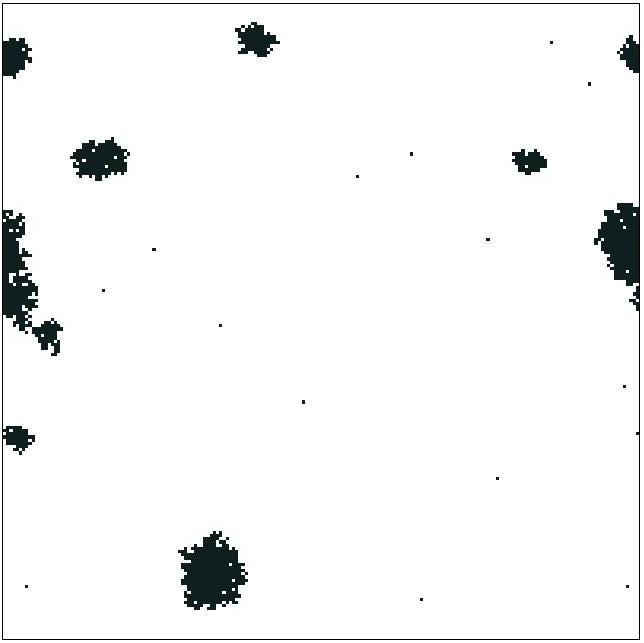}\label{fig:MD10}} &{\includegraphics[width=1.0in]{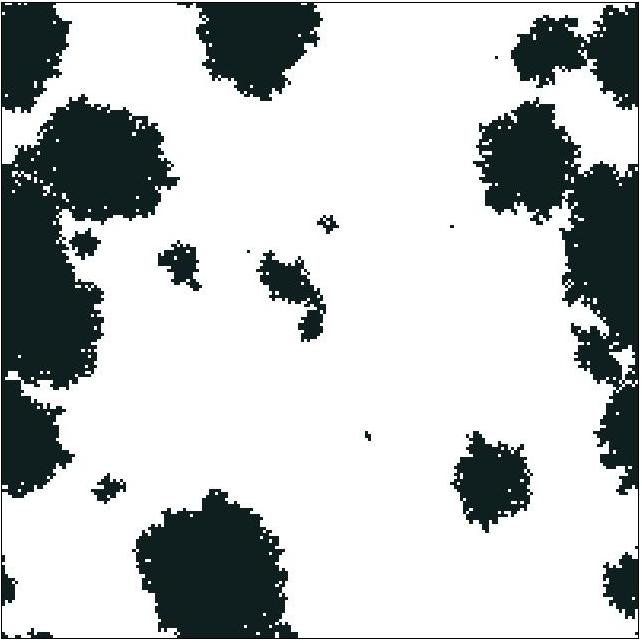}\label{fig:MD20}}
&{\includegraphics[width=1.0in]{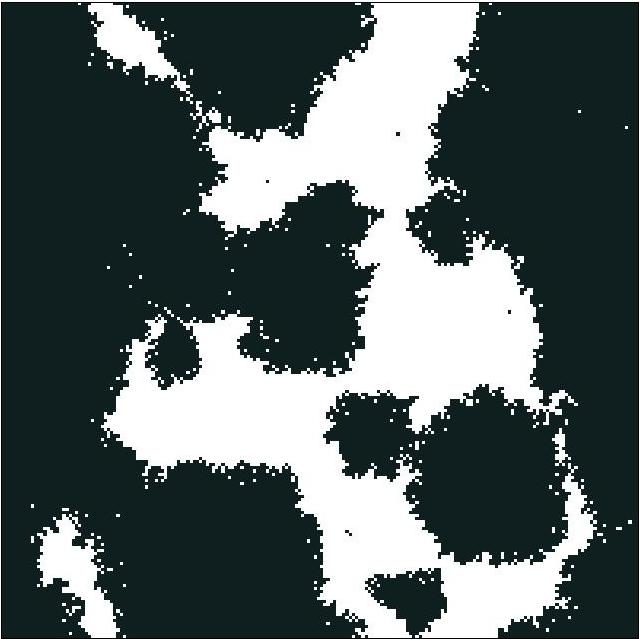}\label{fig:MD30}}\\
{\includegraphics[width=1.0in]{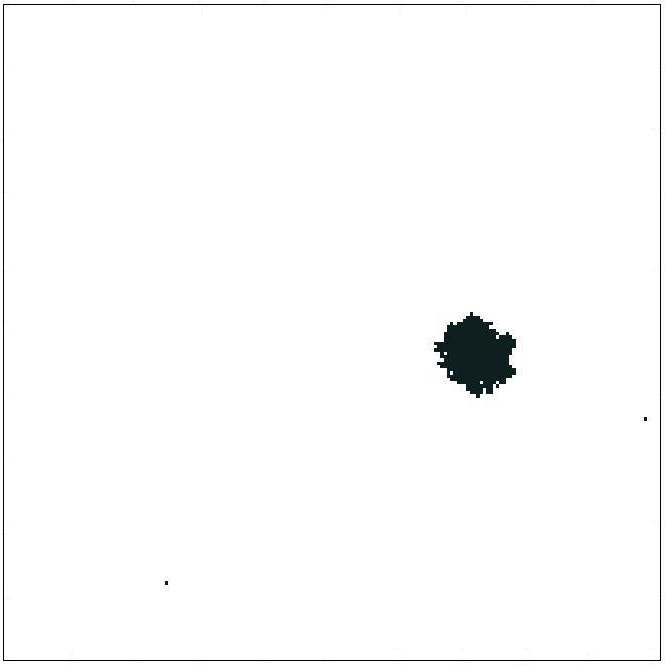}\label{fig:SD20}} &{\includegraphics[width=1.0in]{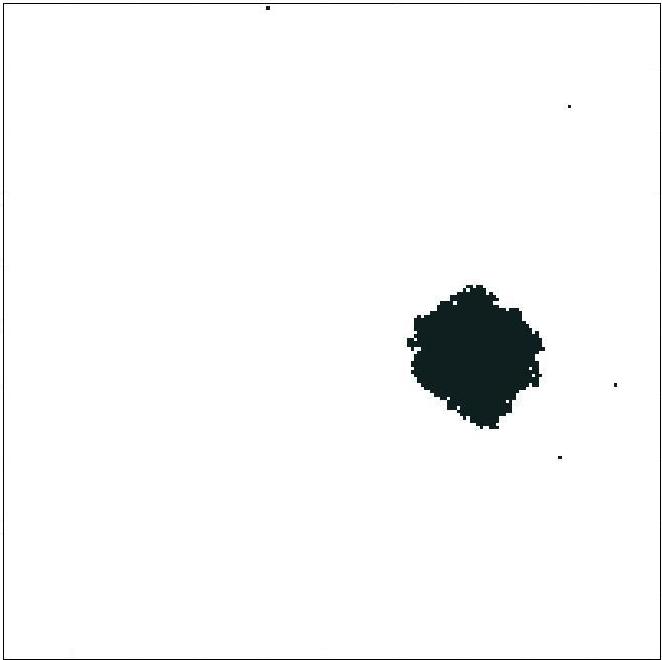}\label{fig:SD30}}&{\includegraphics[width=1.0in]{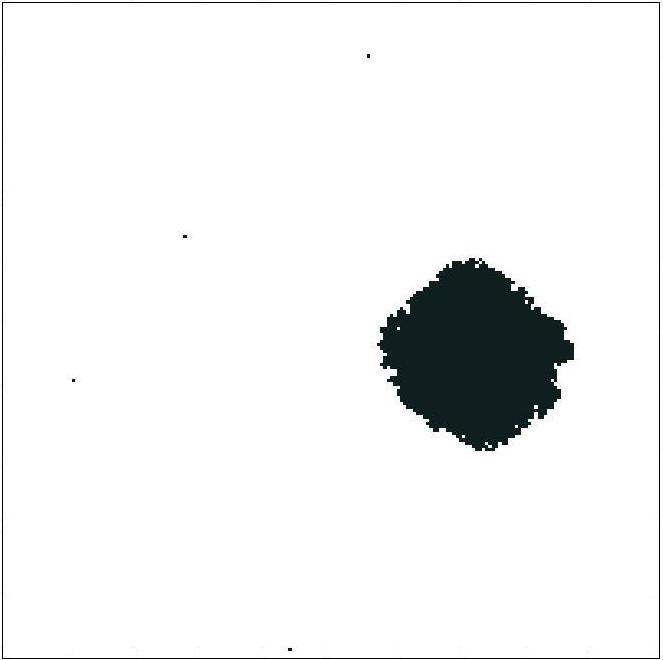}\label{fig:SD40}}\\
\end{array}$
\end{center}
\caption{Tracking the droplets for the NNNKI model at T=0.2T$_{c}$ for L=200 using Metropolis dynamics. The system is initialized with all spins up, s$_{i}=1$. The field was then instantaneously reversed and the relaxation of the system studied. The metastable lifetime is the number of MCSS needed for the system to decay to a net zero magnetized state ($m(t)=0$) from a completely magnetized state ($m(t)=\pm1$). Snapshots of the spin configurations in the SF, MD, and SD region are shown in the three rows respectively. The applied field was h$_{0}$=-6J$_{nn}$ (SF), -4J$_{nn}$ (MD), and -3.25J$_{nn}$ (SD). The first row represents the SF region at t=1, 2,and 3 MCSS. The second row represents MD at t=10, 20, and 30 MCSS. The third row represents SD at t=20, 30, and 40 MCSS. Stable s$_{i}$=-1 (metastable s$_{i}$=+1) spins are represented by black (white). Droplet theory for the NNNKI model holds in the MD and the SD region only.}
\label{fig:9}
\end{figure}
\begin{figure}[h]
  {\label{fig:suscep}
  \includegraphics[width=3.5in]{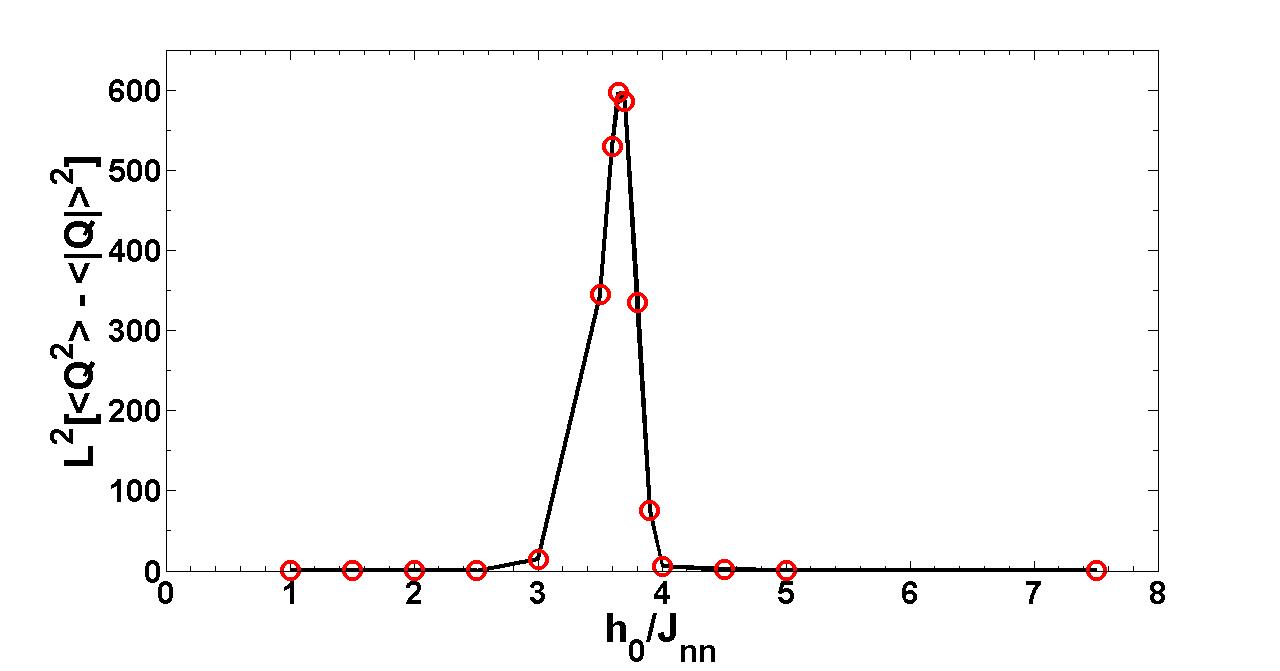}}
  \caption{(Color online) Variation of the order parameter fluctuation X$^{Q}_{L}$=L$^2[\langle Q^{2}\rangle -\langle|Q|^{2}\rangle]$ versus h$_{0}$/J$_{nn}$ at f=10$^{-3}$ and T=0.2T$_{c}$ for L=64 where J$_{nn}$=1. The fluctuation peaks around h$_{0}$/J$_{nn}$=3.65 signifying the DPT transition (see Fig.~\ref{subfig:PQ}). The lines connecting the data points are a guide to the eye.} \label{fig:10}
\end{figure}

The nature of the DPT in the NNKI model was confirmed to be second order by tracking the probablity density of Q ~\cite{Sides:PhysRevE.59.2710}. In this paper we follow the same procedure and compute the P(Q) graphs. The results are shown in Fig.~\ref{subfig:PQ}. There are nine plots which the track the evolution of P(Q) from the SD to the MD and then finally on to the SF region. Each plot is labelled by the letters A - I. The letters are also marked on the DPT plot (see Fig.~\ref{subfig:DPT-NNNKI}). We see that deep in the SD region the probability distribution has its weight shifted to $Q=1$. As the value of h$_0$ is changed some of this weight is transferred to values of Q which are not equal to one. This causes a bimodal distribution to be developed (plot (D) of Fig.~\ref{subfig:PQ}). This bimodal distribution rearranges itself further as we enter the MD region (see plots (E)-(F) of Fig.~\ref{subfig:PQ}). Finally with a further increase in the value of h$_{o}$ the bimodal distribution weight shifts entirely to the Q=0 value with the peak centered around this point (see plots (G)-(I) of Fig.~\ref{subfig:PQ}). This is in the SF region. 

The presence of the bimodal distribution of Q and the observation that P(Q) displays no peak near $Q=0$ in the ordered dynamic phase is a tell-tale sign of the second order phase transition present in the NNNKI system. We conclude from our computations that even though the additional interactions affect the phase boundary, the order of the phase transition is not affected by the presence of these interactions. 

In Fig.~\ref{subfig:Bsurf} we show the density plot for the correlation of the magnetization with the external field for the NNNKI model. From the plot we see that at large magnetic fields and high temperatures the system is maximally correlated. From the DPT plot the parameter regime in which this happens is the SF. The systems magnetization is least correlated with the external field for low fields and low temperatures. From the DPT graph this corresponds to the SD region. In Fig.~\ref{subfig:Asurf} we show the density plot for the hysteresis loop area. The area has its maximum values in the SF region and its minimum values in the MD/SD region. At low temperatures and high field amplitudes the interaction energy dominates the system and the area is a maximum. However, as the temperature is increased thermal effects cause the loop area to diminish but remain finite. The SD region characterized by asymmetric hysteresis loops are the smallest in size and hence the area is small.
\subsection{Hysteresis\label{sec:hysteresis}}
In Fig.~\ref{fig:5} we show the log-log plot of the frequency dispersion of the hysteresis loop area law, A(h$_{o}$,f), for the low frequency range (f $\in[10^{-3},10^{-2}]$) and the SF regime (h$_{o}\geq$ 8) of the NNNKI model. The results are shown for L=64 and L=128 at T=0.2T$_c$. For the NNKI model the low frequency range belongs to f $\leq$10$^{-1}$. Above this range the NNKI model is in the high frequency regime with a different hysteresis loop area law ~\cite{Liu:PhysRevB.65.014416}. Our Monte Carlo simulations of the NNNKI model over the frequency range f $\in[10^{-3},10^{-1}]$ has shown that for f$>$ 10$^{-2}$ the scaling law of the loop area breaks down. Furthermore, hysteresis in the kinetic Ising model is dynamical in origin and disappears in the quasi-static limit. We therefore study the NNNKI model for f $\leq$10$^{-2}$ ~\cite{riknote} and choose the frequency range f $\in[10^{-3},10^{-2}]$.  
\begin{table}[h]
\caption{Hysteresis exponents, a and b, for the loop area law, A(h$_{o}$,f)$\propto$ h$^{a}_{0}$f$^{b}$, for the NNNKI model at a constant temperature. The exponents are computed from the Monte Carlo simulation results and then by a least-squares fitting of the data. The standard deviation for each computed exponent is shown. The results for the NN model are from Ref.~\onlinecite{Chakrabarti:RevModPhys.71.847}.} \begin{ruledtabular}
\begin{tabular*}{\hsize}{c@{\extracolsep{0ptplus1fil}}c@{\extracolsep{0ptplus1fil}}c}
Exponent &NN model &NNN model \\
\colrule
a  & 0.70 & 0.14 $\pm$ 0.01 \\
b  & 0.36 & 0.13 $\pm$ 0.01\\
\end{tabular*}
\end{ruledtabular}
\label{tab:1}
\end{table}

Using the Monte Carlo data we obtain the exponents in the low frequency range of the model. The results are summarized in Table~\ref{tab:1}. We also show the standard deviation for each computed exponent. Since both the couplings are ferromagnetic, the system is aided by the presence of NNN interactions. The presence of NNN interactions in the model introduces additional competition for the external magnetic field to bring about a change. Therefore the response of the system to forming a hysteresis loop of the same area as compared to the NN model is diminished. This is reflected in the power of h$_{0}$ changing from 0.70 (NN) to 0.14 (NNN). The frequency exponent is similarly affected. Furthermore, to confirm that the computed exponents do not suffer from finite-size effects we have studied the lattice size dependence of the hysteresis exponents. We have computed the exponents for L=50, 64, 72, 96, and 128. For these lattice sizes the a-exponents are all $\cong$ 0.14 and the b-exponents are $\cong$ 0.13. The values of the loop area law exponents in this paper are an average of these exponents. 

From Fig.~\ref{fig:5} we observe that the loop area increases as either the frequency or the magnetic field increases. Such a behavior was observed for the NNKI model. Qualitatively both the NNKI and the NNNKI loop areas display similar trends. In Fig.~\ref{fig:6} we show the hysteresis loops corresponding to the external frequencies of Fig.~\ref{fig:5} for L=64, 128 at h$_{0}$=16. We see that as the frequency decreases the loop area also decreases. Both
Figs.~\ref{subfig:Hyst64} and ~\ref{subfig:Hyst128} show similar shapes. In Fig.~\ref{fig:7} we show the scaling of the hysteresis loop area. The two plots display the collapse of the data for either the field amplitude variation (see Fig.~\ref{subfig:hscaling}) or for the frequency variation (see Fig.~\ref{subfig:wscaling}. We observe that for the frequency range and field amplitudes considered the scaling relation for the loop area law is well obeyed. 

\subsection{Metastable lifetime, $\tau(h_{0},T,L,p)$ \label{sec:tau}} 
The key to understanding the observed phenomena in the NNNKI model is to study the metastable lifetime. To determine $\tau$ for this model we performed several instantaneous field reveral simulations ~\cite{Rikvold:PhysRevE.49.5080}. The system was initially prepared in an all up spin configuration. The field was instantaneously reversed and the relaxation of the system then studied. The metastable lifetime is the number of MCSS needed for the system to decay to a net zero magnetized state ($m(t)=0$) from a completely magnetized state ($m(t)=\pm1$). The average metastable lifetime, $\left\langle \tau(h_{o},T,L,p)\right\rangle$, is calculated after 1000 repeated trials. The simulations were performed for L=64 at T=0.8T$_{c}$ and h$_{0}$=0.5J$_{nn}$ for various ratios of the interaction strength p=0,0.5,$\frac{1}{\sqrt{2}}$,1. The results are shown in Fig.~\ref{fig:8}.

From Fig.~\ref{fig:8} we see that the NNNKI model's average $\langle\tau\rangle$ is much greater than the NNKI model's metastable lifetime. Furthermore the lifetime is sensitive to the ratio of the interaction strengths. As mentioned earlier, hysteresis happens due to the competition between the metastable life of the system and the field's period. A change in the value of $\tau$ will therefore cause the system to display a different DPT and hysteretic behavior. In Fig.~\ref{fig:9} we show the formation of the droplets in the SD and MD region to confirm the decay mechanism of the NNNKI model. The SF region is not described by droplets. 
\subsection{Order parameter fluctuation\label{sec:chi}}
For equilibrium systems the fluctuation-dissipation theorem relates the susceptibility to fluctuations in the order parameter. As noted in earlier investigations for the kinetic Ising model it is not obvious what the field conjugate to Q is. Recently, it has been demonstrated that the scaled variance of Q, X$^{Q}_{L}$, can be utilized as a proxy for the susceptibility in the kinetic Ising model to investigate the scaling behavior of the NEQ system near its critical period ~\cite{robb:021124}. The variance of X is defined as\begin{equation}
X^{Q}_{L}=L^{2}[\langle Q^{2}\rangle -\langle|Q|^{2}\rangle]\label{eq:X}\end{equation}In Fig.~\ref{fig:10} we show the results for the fluctuation of Q, X$^{Q}_{L}$ with h$_{0}$/J$_{nn}$ at T=0.2T$_c$ and f=10$^{-3}$ for L=64. The fluctuation peaks around h$_{0}$/J$_{nn}$=3.65 signifying the DPT transition observed in Fig.~\ref{subfig:DPT-NNNKI}. 

\section{Conclusions}\label{sec:summaryconclusion}
In this paper we explore the effects of NNN interaction in the kinetic Ising model. We study the effects on DPT and hysteresis loop area in the low frequency limit of the NNNKI model. We find that including the NNN interaction causes the DPT phase boundary to shift to greater values of magnetic field and temperatures. This can cause the system to undergo an interaction induced DPT. We also confirmed the second order nature of the DPT in the NNNKI model by studying the probability density P(Q) of the dynamic order paramter Q. We find that including the NNN interaction causes the hysteresis loop area law to change from h$^{0.70}_{o}$f$^{0.36}$ (NN) to h$^{0.14\pm 0.01}_{o}$f$^{0.13\pm0.01}$ (NNN). One of our main findings is that the metastable lifetime, $\tau$, which dictates the physics of the NNNKI model is sensitive to additional interactions. Since the presence of additional interactions causes the value of $\tau$ to change it causes the DPT phase boundary and the hysteretic properties to change. Finally, it is our hope that this work will motivate experimentalists to investigate the interaction induced DPT observed in the NNNKI model in real material systems.
\begin{acknowledgements}
WDB and TD thank Tom Colbert, Christian Poppeliers, and Andy Hauger for many helpful and stimulating discussions. WDB thanks Pamplin Student Research Funds (ASU). TD thanks Faculty Research Faculty Development Funds (ASU).   
\end{acknowledgements}
\bibliography{text}

\begin{thebibliography}{37}
\expandafter\ifx\csname natexlab\endcsname\relax\def\natexlab#1{#1}\fi
\expandafter\ifx\csname bibnamefont\endcsname\relax
  \def\bibnamefont#1{#1}\fi
\expandafter\ifx\csname bibfnamefont\endcsname\relax
  \def\bibfnamefont#1{#1}\fi
\expandafter\ifx\csname citenamefont\endcsname\relax
  \def\citenamefont#1{#1}\fi
\expandafter\ifx\csname url\endcsname\relax
  \def\url#1{\texttt{#1}}\fi
\expandafter\ifx\csname urlprefix\endcsname\relax\def\urlprefix{URL }\fi
\providecommand{\bibinfo}[2]{#2}
\providecommand{\eprint}[2][]{\url{#2}}

\bibitem[{\citenamefont{Tom\'e and de~Oliveira}(1990)}]{PhysRevA.41.4251}
\bibinfo{author}{\bibfnamefont{T.}~\bibnamefont{Tom\'e}} \bibnamefont{and}
  \bibinfo{author}{\bibfnamefont{M.~J.} \bibnamefont{de~Oliveira}},
  \bibinfo{journal}{Phys. Rev. A} \textbf{\bibinfo{volume}{41}},
  \bibinfo{pages}{4251} (\bibinfo{year}{1990}).

\bibitem[{\citenamefont{Rao et~al.}(1990)\citenamefont{Rao, Krishnamurthy, and
  Pandit}}]{PhysRevB.42.856}
\bibinfo{author}{\bibfnamefont{M.}~\bibnamefont{Rao}},
  \bibinfo{author}{\bibfnamefont{H.~R.} \bibnamefont{Krishnamurthy}},
  \bibnamefont{and} \bibinfo{author}{\bibfnamefont{R.}~\bibnamefont{Pandit}},
  \bibinfo{journal}{Phys. Rev. B} \textbf{\bibinfo{volume}{42}},
  \bibinfo{pages}{856} (\bibinfo{year}{1990}).

\bibitem[{\citenamefont{Rao and Pandit}(1991)}]{PhysRevB.43.3373}
\bibinfo{author}{\bibfnamefont{M.}~\bibnamefont{Rao}} \bibnamefont{and}
  \bibinfo{author}{\bibfnamefont{R.}~\bibnamefont{Pandit}},
  \bibinfo{journal}{Phys. Rev. B} \textbf{\bibinfo{volume}{43}},
  \bibinfo{pages}{3373} (\bibinfo{year}{1991}).

\bibitem[{\citenamefont{Lo and Pelcovits}(1990)}]{PhysRevA.42.7471}
\bibinfo{author}{\bibfnamefont{W.~S.} \bibnamefont{Lo}} \bibnamefont{and}
  \bibinfo{author}{\bibfnamefont{R.~A.} \bibnamefont{Pelcovits}},
  \bibinfo{journal}{Phys. Rev. A} \textbf{\bibinfo{volume}{42}},
  \bibinfo{pages}{7471} (\bibinfo{year}{1990}).

\bibitem[{\citenamefont{Chakrabarti and
  Acharyya}(1999)}]{Chakrabarti:RevModPhys.71.847}
\bibinfo{author}{\bibfnamefont{B.~K.} \bibnamefont{Chakrabarti}}
  \bibnamefont{and} \bibinfo{author}{\bibfnamefont{M.}~\bibnamefont{Acharyya}},
  \bibinfo{journal}{Rev. Mod. Phys.} \textbf{\bibinfo{volume}{71}},
  \bibinfo{pages}{847} (\bibinfo{year}{1999}).

\bibitem[{\citenamefont{Sides et~al.}(1998{\natexlab{a}})\citenamefont{Sides,
  Rikvold, and Novotny}}]{Sides:PhysRevLett.81.834}
\bibinfo{author}{\bibfnamefont{S.~W.} \bibnamefont{Sides}},
  \bibinfo{author}{\bibfnamefont{P.~A.} \bibnamefont{Rikvold}},
  \bibnamefont{and} \bibinfo{author}{\bibfnamefont{M.~A.}
  \bibnamefont{Novotny}}, \bibinfo{journal}{Phys. Rev. Lett.}
  \textbf{\bibinfo{volume}{81}}, \bibinfo{pages}{834}
  (\bibinfo{year}{1998}{\natexlab{a}}).

\bibitem[{\citenamefont{Acharyya}(1997{\natexlab{a}})}]{PhysRevE.56.2407}
\bibinfo{author}{\bibfnamefont{M.}~\bibnamefont{Acharyya}},
  \bibinfo{journal}{Phys. Rev. E} \textbf{\bibinfo{volume}{56}},
  \bibinfo{pages}{2407} (\bibinfo{year}{1997}{\natexlab{a}}).

\bibitem[{\citenamefont{Acharyya}(1997{\natexlab{b}})}]{PhysRevE.56.1234}
\bibinfo{author}{\bibfnamefont{M.}~\bibnamefont{Acharyya}},
  \bibinfo{journal}{Phys. Rev. E} \textbf{\bibinfo{volume}{56}},
  \bibinfo{pages}{1234} (\bibinfo{year}{1997}{\natexlab{b}}).

\bibitem[{\citenamefont{Acharyya}(1998{\natexlab{a}})}]{PhysRevE.58.174}
\bibinfo{author}{\bibfnamefont{M.}~\bibnamefont{Acharyya}},
  \bibinfo{journal}{Phys. Rev. E} \textbf{\bibinfo{volume}{58}},
  \bibinfo{pages}{174} (\bibinfo{year}{1998}{\natexlab{a}}).

\bibitem[{\citenamefont{Acharyya}(1998{\natexlab{b}})}]{PhysRevE.58.179}
\bibinfo{author}{\bibfnamefont{M.}~\bibnamefont{Acharyya}},
  \bibinfo{journal}{Phys. Rev. E} \textbf{\bibinfo{volume}{58}},
  \bibinfo{pages}{179} (\bibinfo{year}{1998}{\natexlab{b}}).

\bibitem[{\citenamefont{Acharyya}(1999)}]{PhysRevE.59.218}
\bibinfo{author}{\bibfnamefont{M.}~\bibnamefont{Acharyya}},
  \bibinfo{journal}{Phys. Rev. E} \textbf{\bibinfo{volume}{59}},
  \bibinfo{pages}{218} (\bibinfo{year}{1999}).

\bibitem[{\citenamefont{Sides et~al.}(1999)\citenamefont{Sides, Rikvold, and
  Novotny}}]{Sides:PhysRevE.59.2710}
\bibinfo{author}{\bibfnamefont{S.~W.} \bibnamefont{Sides}},
  \bibinfo{author}{\bibfnamefont{P.~A.} \bibnamefont{Rikvold}},
  \bibnamefont{and} \bibinfo{author}{\bibfnamefont{M.~A.}
  \bibnamefont{Novotny}}, \bibinfo{journal}{Phys. Rev. E}
  \textbf{\bibinfo{volume}{59}}, \bibinfo{pages}{2710} (\bibinfo{year}{1999}).

\bibitem[{\citenamefont{Korniss et~al.}(2000)\citenamefont{Korniss, White,
  Rikvold, and Novotny}}]{Korniss:PhysRevE.63.016120}
\bibinfo{author}{\bibfnamefont{G.}~\bibnamefont{Korniss}},
  \bibinfo{author}{\bibfnamefont{C.~J.} \bibnamefont{White}},
  \bibinfo{author}{\bibfnamefont{P.~A.} \bibnamefont{Rikvold}},
  \bibnamefont{and} \bibinfo{author}{\bibfnamefont{M.~A.}
  \bibnamefont{Novotny}}, \bibinfo{journal}{Phys. Rev. E}
  \textbf{\bibinfo{volume}{63}}, \bibinfo{pages}{016120}
  (\bibinfo{year}{2000}).

\bibitem[{\citenamefont{Korniss et~al.}(2002)\citenamefont{Korniss, Rikvold,
  and Novotny}}]{Korniss:PhysRevE.66.056127}
\bibinfo{author}{\bibfnamefont{G.}~\bibnamefont{Korniss}},
  \bibinfo{author}{\bibfnamefont{P.~A.} \bibnamefont{Rikvold}},
  \bibnamefont{and} \bibinfo{author}{\bibfnamefont{M.~A.}
  \bibnamefont{Novotny}}, \bibinfo{journal}{Phys. Rev. E}
  \textbf{\bibinfo{volume}{66}}, \bibinfo{pages}{056127}
  (\bibinfo{year}{2002}).

\bibitem[{\citenamefont{Zhu et~al.}(2004)\citenamefont{Zhu, Dong, and
  Liu}}]{Liu:PhysRevB.70.132403}
\bibinfo{author}{\bibfnamefont{H.}~\bibnamefont{Zhu}},
  \bibinfo{author}{\bibfnamefont{S.}~\bibnamefont{Dong}}, \bibnamefont{and}
  \bibinfo{author}{\bibfnamefont{J.-M.} \bibnamefont{Liu}},
  \bibinfo{journal}{Phys. Rev. B} \textbf{\bibinfo{volume}{70}},
  \bibinfo{pages}{132403} (\bibinfo{year}{2004}).

\bibitem[{\citenamefont{Liu et~al.}(2001)\citenamefont{Liu, Chan, Choy, and
  Ong}}]{Liu:PhysRevB.65.014416}
\bibinfo{author}{\bibfnamefont{J.-M.} \bibnamefont{Liu}},
  \bibinfo{author}{\bibfnamefont{H.~L.~W.} \bibnamefont{Chan}},
  \bibinfo{author}{\bibfnamefont{C.~L.} \bibnamefont{Choy}}, \bibnamefont{and}
  \bibinfo{author}{\bibfnamefont{C.~K.} \bibnamefont{Ong}},
  \bibinfo{journal}{Phys. Rev. B} \textbf{\bibinfo{volume}{65}},
  \bibinfo{pages}{014416} (\bibinfo{year}{2001}).

\bibitem[{\citenamefont{Luse and Zangwill}(1994)}]{Luse:PhysRevE.50.224}
\bibinfo{author}{\bibfnamefont{C.~N.} \bibnamefont{Luse}} \bibnamefont{and}
  \bibinfo{author}{\bibfnamefont{A.}~\bibnamefont{Zangwill}},
  \bibinfo{journal}{Phys. Rev. E} \textbf{\bibinfo{volume}{50}},
  \bibinfo{pages}{224} (\bibinfo{year}{1994}).

\bibitem[{\citenamefont{He and Wang}(1993)}]{PhysRevLett.70.2336}
\bibinfo{author}{\bibfnamefont{Y.-L.} \bibnamefont{He}} \bibnamefont{and}
  \bibinfo{author}{\bibfnamefont{G.-C.} \bibnamefont{Wang}},
  \bibinfo{journal}{Phys. Rev. Lett.} \textbf{\bibinfo{volume}{70}},
  \bibinfo{pages}{2336} (\bibinfo{year}{1993}).

\bibitem[{\citenamefont{Choi et~al.}(1999)\citenamefont{Choi, Lee, Samad, and
  Bland}}]{PhysRevB.60.11906}
\bibinfo{author}{\bibfnamefont{B.~C.} \bibnamefont{Choi}},
  \bibinfo{author}{\bibfnamefont{W.~Y.} \bibnamefont{Lee}},
  \bibinfo{author}{\bibfnamefont{A.}~\bibnamefont{Samad}}, \bibnamefont{and}
  \bibinfo{author}{\bibfnamefont{J.~A.~C.} \bibnamefont{Bland}},
  \bibinfo{journal}{Phys. Rev. B} \textbf{\bibinfo{volume}{60}},
  \bibinfo{pages}{11906} (\bibinfo{year}{1999}).

\bibitem[{\citenamefont{Jiang et~al.}(1995)\citenamefont{Jiang, Yang, and
  Wang}}]{PhysRevB.52.14911}
\bibinfo{author}{\bibfnamefont{Q.}~\bibnamefont{Jiang}},
  \bibinfo{author}{\bibfnamefont{H.-N.} \bibnamefont{Yang}}, \bibnamefont{and}
  \bibinfo{author}{\bibfnamefont{G.-C.} \bibnamefont{Wang}},
  \bibinfo{journal}{Phys. Rev. B} \textbf{\bibinfo{volume}{52}},
  \bibinfo{pages}{14911} (\bibinfo{year}{1995}).

\bibitem[{\citenamefont{Suen et~al.}(1999)\citenamefont{Suen, Lee, Teeter, and
  Erskine}}]{PhysRevB.59.4249}
\bibinfo{author}{\bibfnamefont{J.-S.} \bibnamefont{Suen}},
  \bibinfo{author}{\bibfnamefont{M.~H.} \bibnamefont{Lee}},
  \bibinfo{author}{\bibfnamefont{G.}~\bibnamefont{Teeter}}, \bibnamefont{and}
  \bibinfo{author}{\bibfnamefont{J.~L.} \bibnamefont{Erskine}},
  \bibinfo{journal}{Phys. Rev. B} \textbf{\bibinfo{volume}{59}},
  \bibinfo{pages}{4249} (\bibinfo{year}{1999}).

\bibitem[{\citenamefont{Lee et~al.}(1999)\citenamefont{Lee, Choi, Xu, and
  Bland}}]{PhysRevB.60.10216}
\bibinfo{author}{\bibfnamefont{W.~Y.} \bibnamefont{Lee}},
  \bibinfo{author}{\bibfnamefont{B.-C.} \bibnamefont{Choi}},
  \bibinfo{author}{\bibfnamefont{Y.~B.} \bibnamefont{Xu}}, \bibnamefont{and}
  \bibinfo{author}{\bibfnamefont{J.~A.~C.} \bibnamefont{Bland}},
  \bibinfo{journal}{Phys. Rev. B} \textbf{\bibinfo{volume}{60}},
  \bibinfo{pages}{10216} (\bibinfo{year}{1999}).

\bibitem[{\citenamefont{Suen and Erskine}(1997)}]{PhysRevLett.78.3567}
\bibinfo{author}{\bibfnamefont{J.-S.} \bibnamefont{Suen}} \bibnamefont{and}
  \bibinfo{author}{\bibfnamefont{J.~L.} \bibnamefont{Erskine}},
  \bibinfo{journal}{Phys. Rev. Lett.} \textbf{\bibinfo{volume}{78}},
  \bibinfo{pages}{3567} (\bibinfo{year}{1997}).

\bibitem[{\citenamefont{Binder and
  Landau}(1980)}]{LandauBinder:PhysRevB.21.1941}
\bibinfo{author}{\bibfnamefont{K.}~\bibnamefont{Binder}} \bibnamefont{and}
  \bibinfo{author}{\bibfnamefont{D.~P.} \bibnamefont{Landau}},
  \bibinfo{journal}{Phys. Rev. B} \textbf{\bibinfo{volume}{21}},
  \bibinfo{pages}{1941} (\bibinfo{year}{1980}).

\bibitem[{\citenamefont{Rikvold et~al.}(1994)\citenamefont{Rikvold, Tomita,
  Miyashita, and Sides}}]{Rikvold:PhysRevE.49.5080}
\bibinfo{author}{\bibfnamefont{P.~A.} \bibnamefont{Rikvold}},
  \bibinfo{author}{\bibfnamefont{H.}~\bibnamefont{Tomita}},
  \bibinfo{author}{\bibfnamefont{S.}~\bibnamefont{Miyashita}},
  \bibnamefont{and} \bibinfo{author}{\bibfnamefont{S.~W.} \bibnamefont{Sides}},
  \bibinfo{journal}{Phys. Rev. E} \textbf{\bibinfo{volume}{49}},
  \bibinfo{pages}{5080} (\bibinfo{year}{1994}).

\bibitem[{\citenamefont{Dahmen and Sethna}(1993)}]{PhysRevLett.71.3222}
\bibinfo{author}{\bibfnamefont{K.}~\bibnamefont{Dahmen}} \bibnamefont{and}
  \bibinfo{author}{\bibfnamefont{J.~P.} \bibnamefont{Sethna}},
  \bibinfo{journal}{Phys. Rev. Lett.} \textbf{\bibinfo{volume}{71}},
  \bibinfo{pages}{3222} (\bibinfo{year}{1993}).

\bibitem[{\citenamefont{Sethna et~al.}(1993)\citenamefont{Sethna, Dahmen,
  Kartha, Krumhansl, Roberts, and Shore}}]{PhysRevLett.70.3347}
\bibinfo{author}{\bibfnamefont{J.~P.} \bibnamefont{Sethna}},
  \bibinfo{author}{\bibfnamefont{K.}~\bibnamefont{Dahmen}},
  \bibinfo{author}{\bibfnamefont{S.}~\bibnamefont{Kartha}},
  \bibinfo{author}{\bibfnamefont{J.~A.} \bibnamefont{Krumhansl}},
  \bibinfo{author}{\bibfnamefont{B.~W.} \bibnamefont{Roberts}},
  \bibnamefont{and} \bibinfo{author}{\bibfnamefont{J.~D.} \bibnamefont{Shore}},
  \bibinfo{journal}{Phys. Rev. Lett.} \textbf{\bibinfo{volume}{70}},
  \bibinfo{pages}{3347} (\bibinfo{year}{1993}).

\bibitem[{\citenamefont{Sides et~al.}(1998{\natexlab{b}})\citenamefont{Sides,
  Rikvold, and Novotny}}]{Sides:PhysRevE.57.6512}
\bibinfo{author}{\bibfnamefont{S.~W.} \bibnamefont{Sides}},
  \bibinfo{author}{\bibfnamefont{P.~A.} \bibnamefont{Rikvold}},
  \bibnamefont{and} \bibinfo{author}{\bibfnamefont{M.~A.}
  \bibnamefont{Novotny}}, \bibinfo{journal}{Phys. Rev. E}
  \textbf{\bibinfo{volume}{57}}, \bibinfo{pages}{6512}
  (\bibinfo{year}{1998}{\natexlab{b}}).

\bibitem[{\citenamefont{Buend\'{\i}a and Rikvold}(2008)}]{buenda:051108}
\bibinfo{author}{\bibfnamefont{G.~M.} \bibnamefont{Buend\'{\i}a}}
  \bibnamefont{and} \bibinfo{author}{\bibfnamefont{P.~A.}
  \bibnamefont{Rikvold}}, \bibinfo{journal}{Phys. Rev. E}
  \textbf{\bibinfo{volume}{78}}, \bibinfo{pages}{051108}
  (\bibinfo{year}{2008}).

\bibitem[{\citenamefont{Park et~al.}(2004)\citenamefont{Park, Rikvold,
  Buend\'ia, and Novotny}}]{PhysRevLett.92.015701}
\bibinfo{author}{\bibfnamefont{K.}~\bibnamefont{Park}},
  \bibinfo{author}{\bibfnamefont{P.~A.} \bibnamefont{Rikvold}},
  \bibinfo{author}{\bibfnamefont{G.~M.} \bibnamefont{Buend\'ia}},
  \bibnamefont{and} \bibinfo{author}{\bibfnamefont{M.~A.}
  \bibnamefont{Novotny}}, \bibinfo{journal}{Phys. Rev. Lett.}
  \textbf{\bibinfo{volume}{92}}, \bibinfo{pages}{015701}
  (\bibinfo{year}{2004}).

\bibitem[{\citenamefont{Robb et~al.}(2008)\citenamefont{Robb, Xu, Hellwig,
  McCord, Berger, Novotny, and Rikvold}}]{robb134422}
\bibinfo{author}{\bibfnamefont{D.~T.} \bibnamefont{Robb}},
  \bibinfo{author}{\bibfnamefont{Y.~H.} \bibnamefont{Xu}},
  \bibinfo{author}{\bibfnamefont{O.}~\bibnamefont{Hellwig}},
  \bibinfo{author}{\bibfnamefont{J.}~\bibnamefont{McCord}},
  \bibinfo{author}{\bibfnamefont{A.}~\bibnamefont{Berger}},
  \bibinfo{author}{\bibfnamefont{M.~A.} \bibnamefont{Novotny}},
  \bibnamefont{and} \bibinfo{author}{\bibfnamefont{P.~A.}
  \bibnamefont{Rikvold}}, \bibinfo{journal}{Phys. Rev. B}
  \textbf{\bibinfo{volume}{78}}, \bibinfo{pages}{134422}
  (\bibinfo{year}{2008}).

\bibitem[{\citenamefont{Landau and Binder}(2000)}]{LandauBinderbook}
\bibinfo{author}{\bibfnamefont{D.~P.} \bibnamefont{Landau}} \bibnamefont{and}
  \bibinfo{author}{\bibfnamefont{K.}~\bibnamefont{Binder}},
  \emph{\bibinfo{title}{A Guide to Monte Carlo Simulations in Statistical
  Physics}} (\bibinfo{publisher}{Cambridge University Press},
  \bibinfo{year}{2000}).

\bibitem[{\citenamefont{Newman and Barkema}(1999)}]{NewmanBarkemabook}
\bibinfo{author}{\bibfnamefont{M.~E.~J.} \bibnamefont{Newman}}
  \bibnamefont{and} \bibinfo{author}{\bibfnamefont{G.~T.}
  \bibnamefont{Barkema}}, \emph{\bibinfo{title}{Monte Carlo Methods in
  Statistical Physics}} (\bibinfo{publisher}{Oxford University Press},
  \bibinfo{year}{1999}).

\bibitem[{\citenamefont{Sethna}(2006)}]{Sethnabook}
\bibinfo{author}{\bibfnamefont{J.~P.} \bibnamefont{Sethna}},
  \emph{\bibinfo{title}{Statistical mechanics: entropy, order parameters, and
  complexity}} (\bibinfo{publisher}{Oxford University Press},
  \bibinfo{year}{2006}).

\bibitem[{\citenamefont{Gould et~al.}(2006)\citenamefont{Gould, Tobochnik, and
  Christian}}]{GouldTobochnikbook}
\bibinfo{author}{\bibfnamefont{H.}~\bibnamefont{Gould}},
  \bibinfo{author}{\bibfnamefont{J.}~\bibnamefont{Tobochnik}},
  \bibnamefont{and}
  \bibinfo{author}{\bibfnamefont{W.}~\bibnamefont{Christian}},
  \emph{\bibinfo{title}{An Introduction to Computer Simulation Methods:
  Applications to Physical Systems}} (\bibinfo{publisher}{Addison Wesley},
  \bibinfo{year}{2006}), \bibinfo{edition}{3rd} ed.

\bibitem[{rik()}]{riknote}
\bibinfo{note}{Rikvold and collaborators have predicted that the extremely low
  frequency dispersion of the NNKI hysteresis loop area law is logarithmic. In
  our computations by choosing f $\in[10^{-3},10^{-2}]$ range we are assuming
  that we are not in such a regime of the NNNKI model.}

\bibitem[{\citenamefont{Robb et~al.}(2007)\citenamefont{Robb, Rikvold, Berger,
  and Novotny}}]{robb:021124}
\bibinfo{author}{\bibfnamefont{D.~T.} \bibnamefont{Robb}},
  \bibinfo{author}{\bibfnamefont{P.~A.} \bibnamefont{Rikvold}},
  \bibinfo{author}{\bibfnamefont{A.}~\bibnamefont{Berger}}, \bibnamefont{and}
  \bibinfo{author}{\bibfnamefont{M.~A.} \bibnamefont{Novotny}},
  \bibinfo{journal}{Phys. Rev. E} \textbf{\bibinfo{volume}{76}},
  \bibinfo{pages}{021124} (\bibinfo{year}{2007}).

\end{thebibliography}
\end{document}